\documentclass[fleqn,usenatbib]{mnras}

\usepackage[T1]{fontenc}

\DeclareRobustCommand{\VAN}[3]{#2}
\let\VANthebibliography\thebibliography
\def\thebibliography{\DeclareRobustCommand{\VAN}[3]{##3}\VANthebibliography}

\usepackage{graphicx}	
\usepackage{amsmath}	
\usepackage{amssymb}	
\usepackage{multicol}        
\usepackage{bm}		
\usepackage{pdflscape}	
\usepackage{makecell}
\usepackage{multirow}
\usepackage[normalem]{ulem}
\usepackage{orcidlink}

\usepackage{subcaption}
\usepackage{caption}

\usepackage{newtxtext,newtxmath}
\usepackage{xcolor}

\usepackage{newtxtext,newtxmath}

\newcommand{\hmpc}{$h^{-1}$Mpc}
\newcommand{\ahod}{\textsc{AbacusHOD}}
\newcommand{\tng}{\textsc{IllustrisTNG}}

\renewcommand{\arraystretch}{1.3}

\title[ELG Conformity]{Unraveling emission line galaxy conformity at $z \sim 1$ with DESI early data}

\author[Yuan et al.]{\parbox[t]{0.9\textwidth}{\vspace{-0.7cm}
Sihan Yuan$^{1,2}$\thanks{E-mail: sihany@stanford.edu}\orcidlink{0000-0002-5992-7586},
Risa H. Wechsler$^{1,2, 3}$\orcidlink{0000-0003-2229-011X},
Yunchong Wang$^{1}$,\orcidlink{0000-0001-8913-626X}
Mithi A.C. de los Reyes$^{1,4}$\orcidlink{0000-0002-4739-046X},
Justin Myles$^{1,5}$,
Antoine Rocher$^{6}$\orcidlink{0000-0003-4349-6424},
Boryana Hadzhiyska$^{8}$\orcidlink{0000-0002-2312-3121},
Jessica Nicole Aguilar$^{8}$,
Steven Ahlen$^{9}$\orcidlink{0000-0001-6098-7247},
David Brooks$^{11}$,
Todd Claybaugh$^{8}$, 
Shaun Cole$^{12}$\orcidlink{0000-0002-5954-7903},
Axel de la Macorra$^{13}$\orcidlink{0000-0002-1769-1640},
Jaime E. Forero-Romero$^{14}$\orcidlink{0000-0002-2890-3725},
Satya Gontcho A Gontcho$^{8}$\orcidlink{0000-0003-3142-233X}, 
Julien Guy$^{8}$, 
Klaus Honscheid$^{15,16}$,
Theodore Kisner$^{8}$\orcidlink{0000-0003-3510-7134}, 
Michael Levi$^{8}$\orcidlink{0000-0003-1887-1018}, 
Marc Manera$^{17}$\orcidlink{0000-0003-4962-8934},
Aaron Meisner$^{18}$\orcidlink{0000-0002-1125-7384},
Ramon Miquel$^{17,19}$,
John Moustakas$^{20}$\orcidlink{0000-0002-2733-4559},
Jundan Nie$^{21}$\orcidlink{0000-0001-6590-8122},
Nathalie Palanque-Delabrouille$^{8, 28}$\orcidlink{0000-0003-3188-784X}, 
Claire Poppett$^{8,22}$,
Mehdi Rezaie$^{23}$\orcidlink{0000-0001-5589-7116},
Ashley J. Ross$^{16}$, 
Graziano Rossi$^{24}$,
Eusebio Sanchez$^{25}$\orcidlink{0000-0002-9646-8198},
Michael Schubnell$^{26}$,
Hee-Jong Seo$^{27}$\orcidlink{0000-0002-6588-3508},
Gregory Tarlé$^{26}$\orcidlink{0000-0003-1704-0781},
Benjamin Alan Weaver$^{18}$,
and Zhimin Zhou$^{21}$\orcidlink{0000-0002-4135-0977}
}
\vspace{0.3cm}
\\
\parbox{\textwidth}{
The authors' affiliations are shown in Appendix \ref{sec:affiliations}}.
\vspace{-0.5cm}}
\date{Accepted XXX. Received YYY; in original form ZZZ}

\pubyear{2015}

\begin{document}
\label{firstpage}
\pagerange{\pageref{firstpage}--\pageref{lastpage}}
\maketitle

\begin{abstract}
Emission line galaxies (ELGs) are now the preeminent tracers of large-scale structure at $z > 0.8$ due to their high density and strong emission lines, which enable accurate redshift measurements. However, relatively little is known about ELG evolution and the ELG--halo connection, exposing us to potential modeling systematics in cosmology inference using these sources. 
In this paper, we propose a physical picture of ELGs and improve ELG--halo connection modeling using a variety of observations and simulated galaxy models. We investigate DESI-selected ELGs in COSMOS data, and infer that ELGs are rapidly star-forming galaxies with a large fraction exhibiting disturbed morphology, implying that many of them are likely to be merger-driven starbursts. 
We further postulate that the tidal interactions from mergers lead to correlated star formation in central--satellite ELG pairs, a phenomenon dubbed ``conformity.'' We argue for the need to include conformity in the ELG--halo connection using galaxy models such as \tng, and by combining observations such as the DESI ELG auto-correlation, ELG cross-correlation with Luminous Red Galaxies (LRGs), and ELG--cluster cross-correlation. We also explore the origin of conformity using the UniverseMachine model and elucidate the difference between conformity and the well-known galaxy assembly bias effect. 

\end{abstract}




\section{Introduction}

In the $\Lambda$CDM cosmology framework, galaxies live in clumps of dark matter called dark matter halos \citep{1978White, 2002Cooray}. By studying the galaxy--halo connection, we can succinctly link the observed galaxy field to the dark matter density field calculated from cosmological theory (see \citealt{2018Wechsler} for a review). One popular framework for empirically connecting galaxies to halos through a set of simple yet flexible probabilistic models is known as the Halo Occupation Distribution \citep[HOD; e.g.][]{1998Jing, 2000Peacock, 2001Scoccimarro, 2001White, 2002Berlind, 2003Berlind, 2005Zheng, 2007bZheng}.

The HOD has been highly successful in describing the non-linear scale clustering of magnitude-limited galaxy samples in past galaxy redshift surveys \citep[e.g.][]{2011Zehavi, 2013Parejko, 2014Guo, 2015cGuo, 2016Rodriguez, 2020Alam, 2020Avila, 2021bYuan}. Such HOD studies not only produce models and mocks that match the observed clustering in great detail \citep[e.g.][]{2020Smith, 2021Rossi, 2021Alam}, but can reveal aspects of galaxy evolution physics \citep[e.g.][]{2019bLange, 2020Alam, 2021Yuan, 2022Wang, 2022Linke}. Most recently, simulation-based forward models has employed HODs to derive cosmology constraints from non-linear scales \citep[e.g.][]{2021Lange, 2021Kobayashi, 2021Chapman, 2022bYuan, 2022Zhai}. 

The need to understand the galaxy--halo connection is especially critical given the precision of the next generation of large-scale structure surveys, including the ongoing the Dark Energy Spectroscopic Instrument \citep[DESI;][]{2013Levi,2016DESI}.
DESI will obtain spectroscopic measurements of 40 million galaxies and quasars in a 14,000 deg$^2$ footprint. This represents an order-of-magnitude improvement both in the volume surveyed and the number of galaxies measured over previous surveys. 
DESI will achieve this by extending to significantly higher redshift, relying on a dense sample of ELGs in $0.8 < z < 1.6$. ELGs are ideal tracers because of their high density and because they exhibit a strong characteristic [O\,II] doublet emission line, which allows for optimal spectroscopic redshift measurements of $z=$ 0.5-2 galaxies \citep{2010Drinkwater, 2017Raichoor}. For this reason, ELGs have become the premier galaxy tracers at $z > 1$ not just for DESI, but also for a slew of next-gen cosmological surveys including Prime Focus Spectrograph \citep{2014Takada} and the Roman Space Telescope \citep{2013Spergel}.  
However, despite their importance in realizing the goals of next generation of cosmological surveys, relatively little is known about ELGs and ELG--halo connection. Moreover, ELG properties might be sensitive to redshift evolution and the details of their selection. Thus, there exists an urgent need to study their physical nature and build robust models for ELG--halo connection. 

Several studies have modeled the ELG--halo connection using the relatively small eBOSS \citep{2016Dawson} ELG sample \citep{2016Favole, 2019Guo, 2020Avila}. They found that the ELG clustering can be well explained by an HOD model where the central occupation peaks at approximately $10^{12}h^{-1}M_\odot$ halos and the satellite occupation follows a power law. More recent studies have started to find deviations from this vanilla model. \cite{2020Alam} found a small discrepancy between such vanilla HOD model and data in the ELG$\times$LRG cross-correlations. The authors interpreted the discrepancy as potential evidence for correlated quenching, which converts some ELG satellites near LRG centrals to LRGs. Simultaneously, studies have used hydrodynamical models to investigate the ELG--halo connection. For example, \cite{2022mHadzhiyska} found evidence for spatially correlated star formation in ELGs and a need to extend the vanilla HOD model for ELGs.  

Spatial correlation in galaxy properties such as star formation rate and colors is broadly referred to as galactic conformity. The term was first coined in the context of large galaxy surveys in \cite{2006Weinmann}, and the effect was confirmed observationally in numerous subsequent studies \citep[e.g.][]{2008Ann, 2011Prescott, 2012Wang, 2013Kauffmann, 2015Hartley, 2015Hearin, 2015Knobel, 2017Pahwa, 2018Treyer, 2023Ayromlou}. Galactic conformity is also readily produced in hydrodynamical simulations of galaxy formation \citep[e.g.g][]{2016Bray, 2018Rafieferantsoa}. However, it is not immediately clear what is driving the observed galactic conformity. While it is widely agreed on that galaxies' dependencies on secondary halo properties other than mass, i.e. galaxy assembly bias, can produce galactic conformity, other effects that associate galaxy evolution with environmental effects can also induce galactic conformity \citep[e.g.][]{2015Paranjape, 2018Zu, 2018Tinker}. 

With DESI One-Percent Data \citep{edr, sv}, we have for the first time measured ELG clustering at percent-level accuracy deep into 1-halo scales. \cite{2023Rocher} analyzed the clustering of the DESI One-Percent Survey ELG sample in an extended HOD framework. Surprisingly, the ELG auto-correlation function exhibits a large upturn at $\sim 0.1 h^{-1}$Mpc that is not readily predicted by a vanilla HOD with values calibrated on hydrodynamical simulations or eBOSS ELG HOD fits \citep{2020Alam, 2020Avila}. However, the paper suggested that the small-scale amplitude can be explained by galaxy conformity, which generates excess small-separation pairs by populating ELG satellites in the same halos as ELG centrals. However, the paper also finds a degeneracy between the conformity HOD model and an extreme vanilla HOD model. A vanilla ELG HOD model can generate excess clustering on 1-halo scales by adopting a very small, even negative, satellite occupation scaling index $\alpha$, which ``over-load'' low mass halos with ELG satellites, thus generating a large number of close ELG pairs. \cite{2023Rocher} argued qualitatively that this scenario is physically implausible, but additional arguments are needed to break the conformity-$\alpha$ degeneracy. 

In this paper, we present a physical picture for DESI ELGs that naturally motivates the inclusion of 1-halo conformity and galaxy assembly bias in the ELG HOD model. We achieve this by combining a suite of simulated galaxy models and observations, including the auto and cross-correlation between ELGs and LRGs in the DESI One-Percent Survey. We start by presenting the DESI data sets and the clustering measurements in section~\ref{sec:data}. We use COSMOS data to put forth a physical picture of DESI ELGs in section~\ref{sec:cosmos}. Then, we describe the HOD framework in section~\ref{sec:hod}, and present evidence for ELG conformity and assembly bias in \tng\ in section~\ref{sec:tng}. Taking inspiration from \tng, we construct extended ELG HOD models and apply them to the DESI auto and cross clustering in section~\ref{sec:desi_method} and \ref{sec:desi}. We further address the $\alpha$-conformity degeneracy in section~\ref{sec:redmapper} by measuring $\alpha$ in ELG--cluster cross-correlation. We explore the origin of conformity in section~\ref{sec:um} and elucidate the difference between conformity and galaxy assembly bias. In section~\ref{sec:discuss}, we discuss complementary works on ELG conformity and alternative theories of the origin of conformity. Finally, we conclude in section~\ref{sec:conclude}.

Throughout this paper, we adopt the Planck 2018 $\Lambda$CDM cosmology ($\Omega_c h^2 = 0.1200$, $\Omega_b h^2 = 0.02237$, $\sigma_8 = 0.811355$, $n_s = 0.9649$, $h = 0.6736$, $w_0 = -1$, and $w_a = 0$) \citep{2020Planck}. We use $h^{-1}$Mpc units for distances, $h^{-1}M_\odot$ for halo masses, and $M_\odot$ for stellar masses. 

\section{Data}
\label{sec:data}
The primary goal of this paper is to better understand the physical nature of the DESI ELG sample and propose motivated models to describe their clustering. In this section, we introduce the DESI One-Percent Survey, the LRG and ELG samples, and their respective auto and cross-correlation measurements. 

\subsection{DESI One-Percent Survey}
DESI conducted its One Percent Survey as the third and final phase of its Survey Validation (SV) program in April and May of 2021. Observation fields were chosen to be in 20 non-overlapping `rosettes' selected to cover major datasets from other surveys, including the Cosmic Evolution Survey (COSMOS), Hyper Suprime-Cam (HSC), Dark Energy Survey (DES) deep field, Galaxy And Mass Assembly (GAMA), Great Observatories Origins Deep Survey (GOODS), and anticipated deep fields from future Legacy Survey of Space and Time (LSST) and Euclid observations. Each rosette is observed at least 12 times, resulting in over high completeness for all targeted samples. The One-Percent survey observed approximately 90,000 LRGs, 270,000 ELGs, 30,000 QSOs, and 150,000 low-redshift galaxies known as Bright Galaxy Sample (BGS). We refer readers to \cite{sv} and \cite{edr} for more details. The One Percent Survey essentially produces a smaller but more complete preview version of the upcoming DESI main sample, and is ideal for calibrating galaxy--halo connection models and building high-fidelity DESI mocks.

\subsection{DESI LRGs and ELGs}

The Luminous Red Galaxies (LRGs) are commonly used as large-scale structure tracers due to two main advantages: 1) they are bright galaxies with the prominent 4000\AA\ break in their spectra, thus allowing for relatively easy target selection and redshift measurements; and 2) they are highly biased tracers of the large-scale structure, thus yielding a higher S/N per-object for the BAO measurement compared to typical galaxies. The LRG target selection is defined in \citet{2020Zhou, 2022Zhou}.

Emission Line Galaxies (ELG) will be the largest sample within DESI for large-scale structure studies. ELGs are bright star-forming galaxies at $z > 0.8$ and extend out to redshifts of $z < 1.6$. ELGs are particularly suited for spectroscopic surveys as they can be selected by their strong [O~II] doublet emission lines, which allow for precise redshift measurements. The ELG target selection algorithm is detailed in \cite{2020Raichoor, 2023Raichoor}.


\begin{figure}
    \hspace{-0.3cm}
    \includegraphics[width=0.5\textwidth]{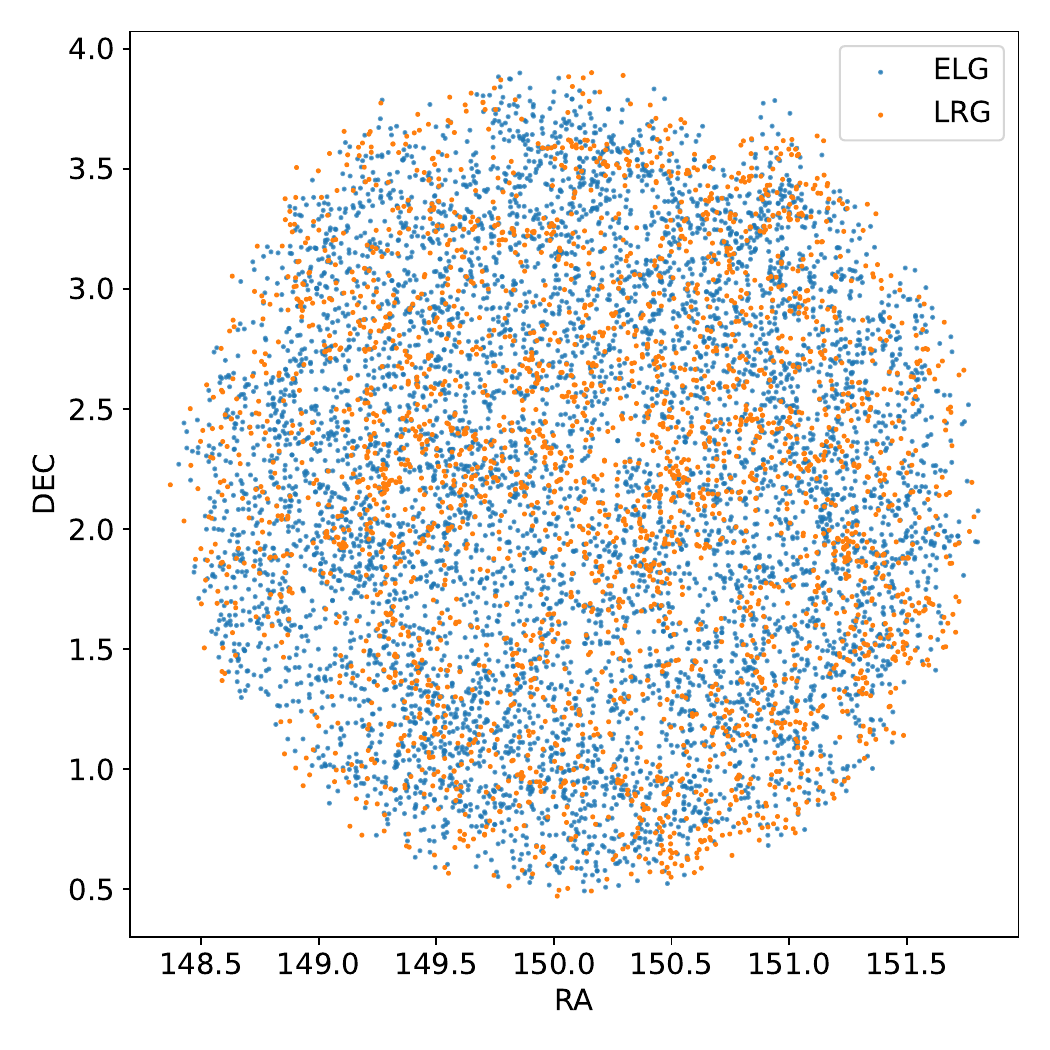}
    \vspace{-0.4cm}
    \caption{Scatter plot showing the projected distribution of ELGs (blue) and LRGs (orange) in $0.8 < z < 1.1$ in one of the 20 rosettes in DESI One-Percent Survey. This specific rosette also covers the COSMOS field. }
    \label{fig:radec}
\end{figure}

\begin{figure}
    \hspace{-0.3cm}
    \includegraphics[width=0.5\textwidth]{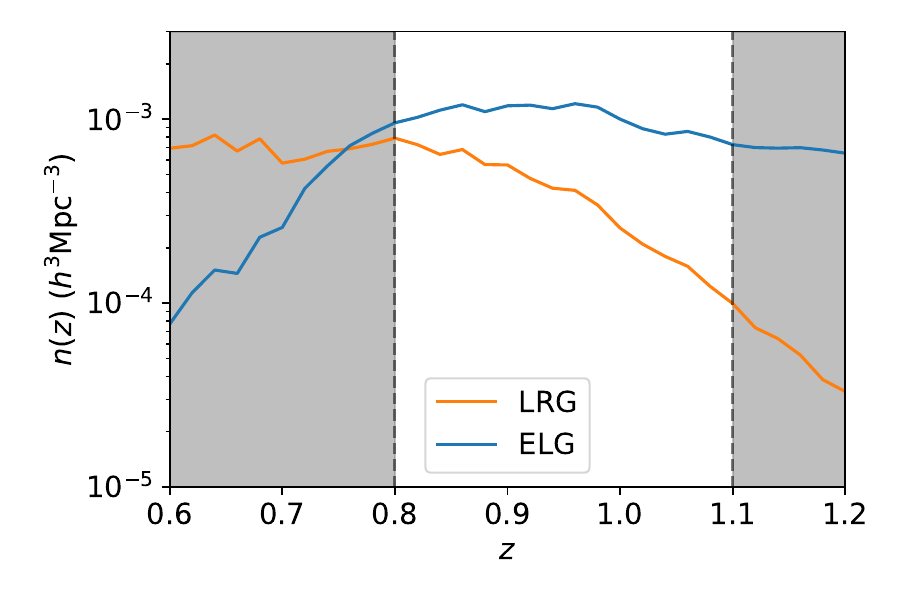}
    \vspace{-0.4cm}
    \caption{The mean number density of the DESI EDR LRG and ELG samples as a function of redshift. For this study, we only consider the two samples in the $0.8 < z < 1.1$ redshift regime, where the ELG sample reaches its peak number density and the LRG sample density drops off.}
    \label{fig:nz}
\end{figure}

Figure~\ref{fig:radec} visualizes the projected distribution of DESI ELGs and LRGs in redshift range $0.8 < z < 1.1$ in one of the 20 rosettes in DESI One-Percent Survey. This rosette specifically covers the COSMOS field. We see strong spatial clustering in both samples. 
Figure~\ref{fig:nz} shows the mean number density of the DESI EDR LRG and ELG samples as a function of redshift. The dashed lines denote the redshift regime we study in this analysis. Within redshift $0.8 < z < 1.1$, we obtain roughly 34,000 LRGs and 125,000 ELGs, for an average number density of $3.8\times 10^{-4}h^3$Mpc$^{-3}$ and $1.0\times 10^{-3}h^3$Mpc$^{-3}$, respectively.

\subsection{Observed clustering}
\begin{figure*}
    \hspace{-0.7cm}
    \includegraphics[width=1\textwidth]{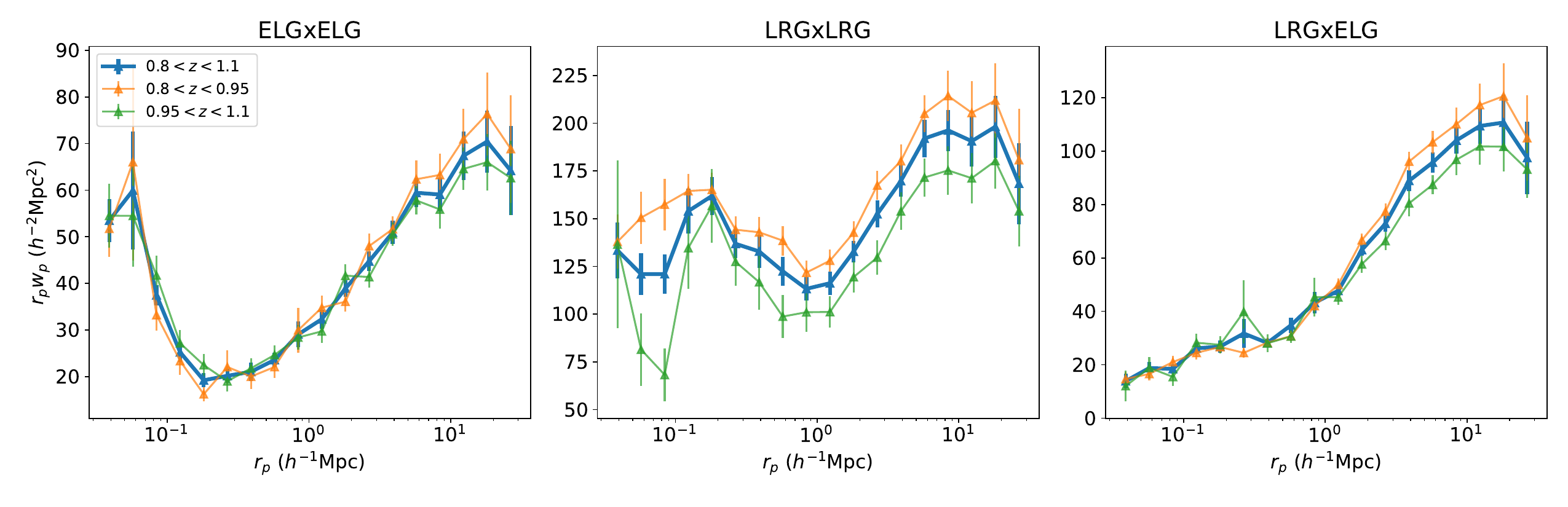}
    \vspace{-0.2cm}
    \caption{The projected auto and cross correlation functions of the DESI One-Percent ELG and LRG samples. The error bars on the DESI samples come from 60 jackknife regions in the survey footprint. The highlighted blue curves show the clustering in $0.8 < z < 1.1$, which we use for the following analysis. The orange and green lines show the clustering in $0.8 < z < 0.95$ and $0.95 < z < 1.1$, respectively. }
    \label{fig:wps}
\end{figure*}
In this paper, we consider the projected correlation functions within and between the DESI EDR LRG and ELG samples. The construction of the EDR catalogs is described in \cite{edr}. Briefly, these LSS catalogs apply quality cuts to the data samples and provide matched random catalogs that trace the angular footprint and $dN/dz$ of the data. \cite{altMTL} describes simulations of the DESI One Percent Survey fiber assignment in order to encode via bits the fiber assignment probability of each target and thus any joint probabilities of observation for a given set of targets.
We use this information to determine the pairwise-inverse-probability \citep{2017Bianchi} weights to use in our clustering measurements. We further apply angular up-weighting (PIP+ANG) \citep{2017Percival}.  \cite{2020Mohammad} showed that this weighting scheme provides an unbiased clustering down to 0.1\hmpc.

The One Percent Survey LSS catalogs also include the so-called `FKP' \citep{FKP} weights in order to properly weight each volume element with respect to how each sample's number density changes with redshift,
\begin{equation}
    w_{\rm FKP} = 1/(1+n(z) P_0)
\end{equation}
where $n(z)$ is the weighted number per volume, and $P_0$ is a fiducial power-spectrum amplitude. For a detailed description of the weights and systematics treatment, we refer the readers to \cite{edr}.

The 2-point correlation function (2PCF) can be computed using the \citet{1993Landy} estimator:
\begin{equation}
    \xi(r_p, r_\pi) = \frac{DD - 2DR + RR}{RR},
    \label{equ:xi_def}
\end{equation}
where $DD$, $DR$, and $RR$ are the normalized numbers of data-data, data-random, and random-random pair counts in each bin of $(r_p, r_\pi)$. $r_p$ and $r_\pi$ are transverse and line-of-sight (LoS) separations in comoving units. We can then compress the full shape $\xi(r_p, r_\pi)$ to the projected galaxy 2PCF $w_p$, which is the line-of-sight integral of $\xi(r_p, r_\pi)$,
\begin{equation}
w_p(r_p) = 2\int_0^{r_{\mathrm{\pi, max}}} \xi(r_p, r_\pi)dr_\pi,
\label{equ:wp_def}
\end{equation}

We show the DESI EDR LRG and ELG auto and cross-correlations with the orange curves in Figure~\ref{fig:wps}. The error bars represent $68\%$ intervals and come from 60 jackknife regions of the One-Percent Survey footprint. The blue curves show the clustering measurement over the entire redshift range of $0.8 < z < 1.1$, whereas the orange and green curves show the clustering in the lower and higher redshift halves. We see that there is insignificant redshift evolution in the ELG auto-correlation function and the ELG$\times$LRG cross-correlation function (see \cite{2023Gao} for a more detailed discussion). There is more redshift evolution in the amplitude of the LRG clustering, at approximately $20\%$ between the two lower and higher redshift halves. However, the significance of the deviation from the mean measurement in blue is rather low at only 1-2$\sigma$. Furthermore, we measured the evolution of the LRG sample in \cite{2023Yuan}, where we found a mild change of $\sim 10\%$ in the LRG linear bias in $0.8 < z < 1.1$.

In this paper, we ignore the redshift evolution of the two samples and assume that the samples can be approximated with a single HOD. This is a reasonable assumption given the mild evolution. Nevertheless, we acknowledge that this is an important caveat that can impact our conclusions. We reserve a more detailed analysis of the redshift evolution of the two samples for future work. 

The ELG auto-correlation function shows a dramatic upturn below $r_p < 0.2h^{-1}$Mpc, whereas the ELG$\times$LRG cross-correlation shows very low amplitude at similar scales. This suggests that ELGs are highly clustered on 1-halo scales, but there are very few ELG-LRG pairs on 1-halo scales. These features were not immediately obvious in eBOSS \citep{2020Avila, 2020Alam}, but have since become a focus of ELG--halo connection modeling efforts in DESI \citep{2023Rocher, gao_conf}. The key objective of this paper is to improve our understanding of the nature of ELGs and the modeling of their clustering.

\section{Probing ELGs in the COSMOS field}
\label{sec:cosmos}
There is a relative dearth of literature exploring the nature of ELGs. In this section, we explore the nature and properties of ELGs using existing data in the COSMOS field.

COSMOS is a deep, wide area, multi-wavelength survey over 2 square degrees aimed at measuring the evolution of galaxies on scales, and is also covered in DESI EDR. The field has been observed at all accessible wavelengths from the X-ray to the radio with most of the major space-based (Hubble, Spitzer, GALEX, XMM, Chandra, Herschel, NuStar) and ground-based telescopes (Keck, Subaru, Very Large Array (VLA), European Southern Observatory Very Large Telescope (ESO-VLT), United Kingdom Infrared Telescope (UKIRT), The National Optical Astronomical Observatory (NOAO) Badde and Blanco telescopes, the Canada France Hawaii Telescope (CFHT), and others). 
In this section, we identify DESI ELGs in the COSMOS2020 catalogs \citep{2021Weaver}. We study the inferred galaxy properties such as stellar mass and star formation rate from existing measurements and also present a preliminary morphology analysis using Hubble Advanced Camera for Surveys (HST ACS) images. 

\begin{figure*}
    \hspace{-0.7cm}
    \includegraphics[width=1\textwidth]{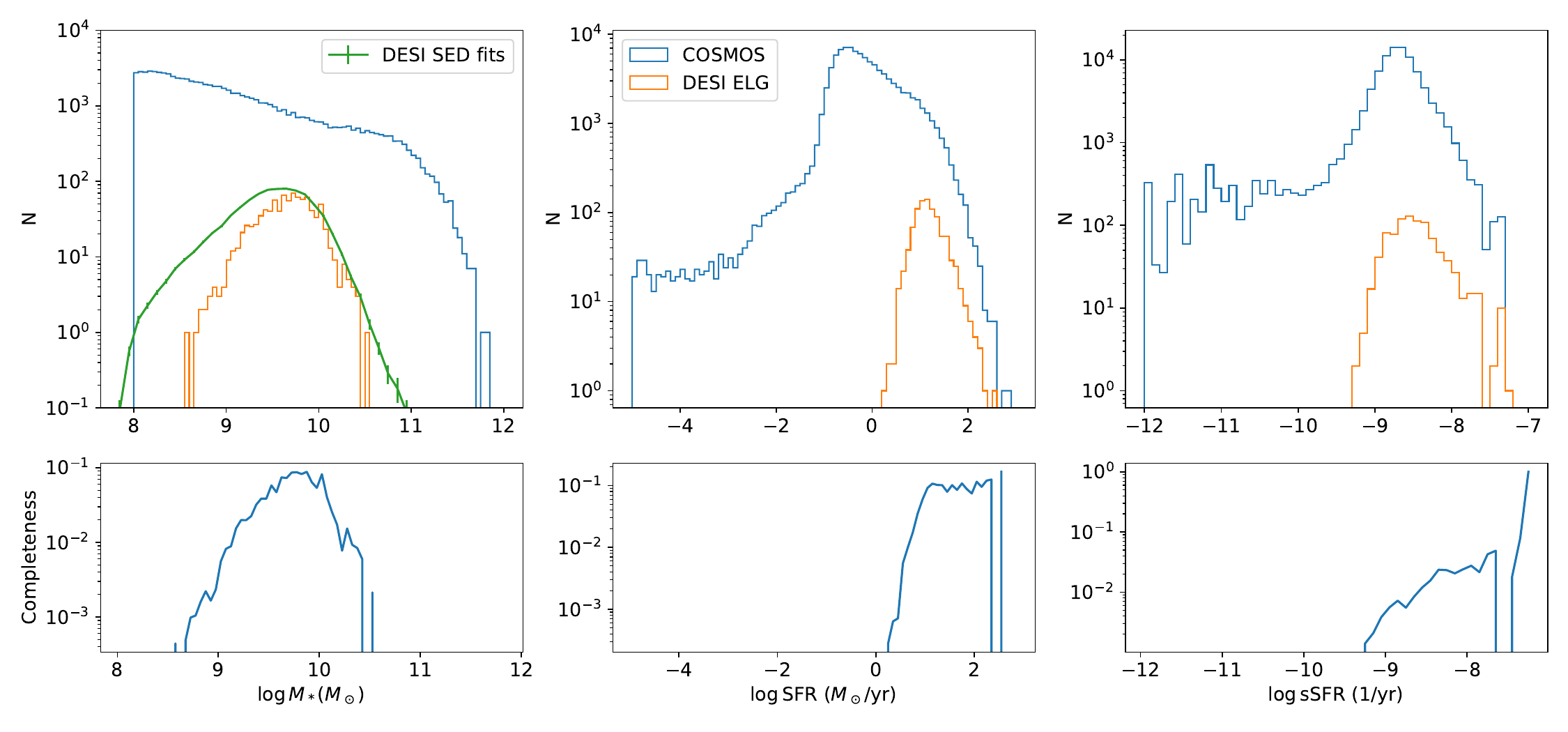}
    \vspace{-0.2cm}
    \caption{The distribution of stellar masses (left panel), star formation rate (middle panel), and specific star formation rate (right panel) of DESI ELGs in COSMOS field compared to all galaxies in COSMOS2020 catalog in $0.8 < z < 1.1$. In green, we also show the stellar mass function of \citet{2023Gao} derived from fitting DESI ELG spectra. The bottom panels show the completeness of the DESI ELG sample relative to the COSMOS2020 sample.}
    \label{fig:cosmos_completeness}
\end{figure*}


\begin{figure}
    \hspace{-0.1cm}
    \includegraphics[width=0.5\textwidth]{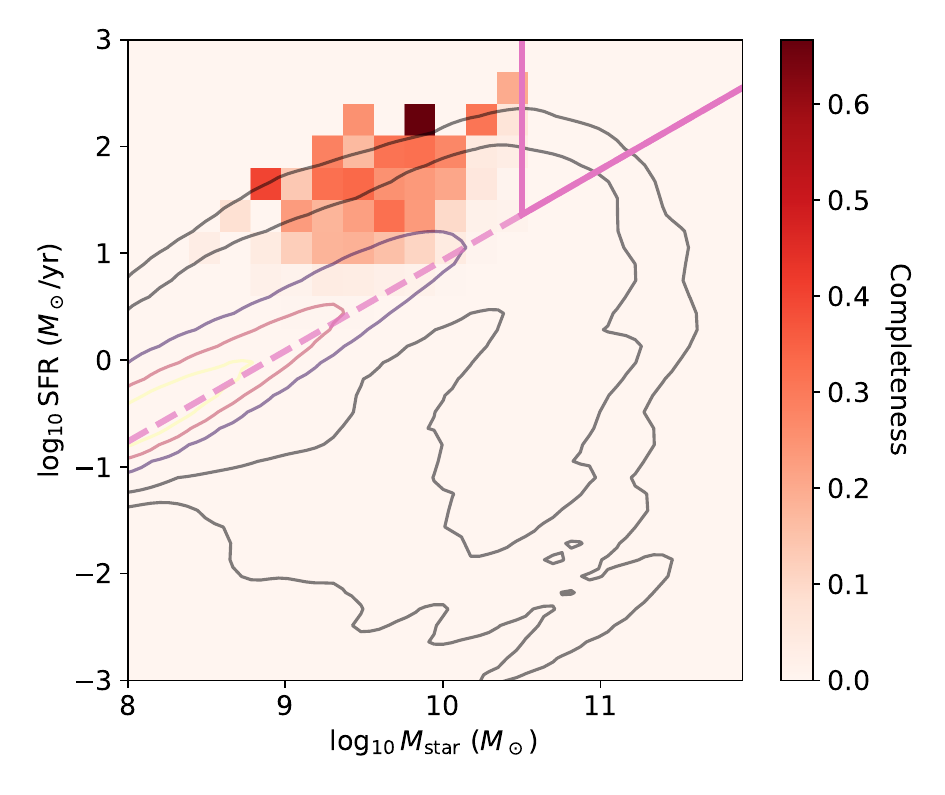}
    \vspace{-0.4cm}
    \caption{The completeness of ELGs relative to the underlying COSMOS galaxies in the stellar mass vs SFR plane. The contours visualize the distribution of 97k COSMOS galaxies in $0.8 < z < 1.1$. The red histograms show the completeness and distribution of ELGs. We plot the dashed line to represent the mean behavior of the star-forming main sequence. The most massive star-forming galaxies selected by the pink box are missing from the ELG sample. We compare them to DESI ELG selection cuts in Figure~\ref{fig:selection}.}
    \label{fig:completeness2d}
\end{figure}
We identify DESI ELGs in the COSMOS2020 catalog by spatially cross-matching with the DESI EDR ELG catalog in $0.8<z<1.1$ with an 1-arcsecond filter. Then amongst the 931 matches, we require that their best-fit photo-$z$ be within 0.1 of the DESI redshift so that their inferred galaxy properties are not significantly biased by their photo-$z$ errors. We identify 929 ELG matches in COSMOS2020.

\subsection{Properties from SED fits}
Figure~\ref{fig:cosmos_completeness} presents the distribution of the inferred stellar masses and star formation rates of the ELG matches in orange, compared to all 97 thousand COSMOS galaxies in $0.8<z<1.1$ in blue. We also compare against the stellar mass function (SMF) of DESI ELGs inferred from $grzW1W2$ 5-band fits \citep[green distribution;][]{2023Gao}. The COSMOS stellar masses and star formation rates were inferred from \textsc{LePhare} \citep{2001Arnouts, 2006Ilbert}, where a SED template library is fit to the observed photometry after fixing the redshift to the photo-$z$ estimated following \cite{2016Laigle}. An uncertainty of 0.2-0.3 dex is expected for the inferred masses due to systematics in the SED fitting method \citep{2019Leja, 2018Sorba}, and potential additional uncertainties may arise from photo-$z$ errors.

The ELG SMF measured via COSMOS cross-matches in orange is consistent with the 5-band Legacy Survey SED fits from \cite{2023Gao} in green above $\log_{10} M_\mathrm{star}/M_\odot \approx 10$. This consistency is remarkable as the two SMFs come from distinct datasets and independent pipelines. The discrepancy at lower masses is possibly due to a combination of photo-$z$ errors in the 5-band fit and other systematics in the SED fitting pipelines at lower masses. Comparing the orange SMF to the full COSMOS SMF in blue, we see that the ELGs are a mass-incomplete sample, with a peak completeness of $\approx 10\%$ at $\log_{10} M_\mathrm{star}/M_\odot \approx 9.8$. As we show later in Section~\ref{sec:desi}, this is consistent with the HOD fits, which show a peak completeness of $\sim 10\%$ in halos of $\sim 10^{12}h^{-1}M_\odot$.

The middle panels of Figure~\ref{fig:cosmos_completeness} show the distribution of ELG star formation rates relative to the full COSMOS2020 sample. Clearly, the DESI ELGs are some of the most actively star-forming galaxies. However, the peak completeness in terms of SFR is also only $\sim 10\%$ at the high SFR end, and the completeness drops off very quickly at $\log_{10}$SFR$<1M_\odot$/yr. A similar trend is seen in the distribution of specific star formation rate (sSFR) in the right panels, where the ELGs represent an incomplete sample of galaxies with the highest sSFRs.


To better understand ELG completeness as a function of physical properties, Figure~\ref{fig:completeness2d} shows ELG completeness in the 2D stellar mass-SFR plane. The 2D contours show the distribution of all galaxies in the COSMOS catalog. We see a star-forming main sequence (SFMS), where we have added a pink dashed line to highlight the mean trend. Note that the pink dashed line is not a fit, but simply serves as visual guide. There is also a quenched sequence to the bottom right of the plot. The completeness of the ELG population is shown with the red histograms. The ELGs are all significantly above the SFMS (by $>1$~dex in most cases), populating a regime that is known as starburst galaxies \citep[e.g.][]{2011Rodighiero, 2015Schreiber, 2018Elbaz}. Several studies have found evidence that starburst galaxies are dominated by merging or interacting galaxy systems \citep{1988Sanders, 2009Urrutia, 2021Moreno}, thus raising the possibility that ELGs preferentially select galaxies undergoing mergers and tidal interactions with nearby galaxies. This merger hypothesis is also consistent with the fact that the ELG completeness increases with SFR. We discuss this further in Section~\ref{sec:morph}.

\begin{figure}
    \hspace{-0.3cm}
    \includegraphics[width=0.5\textwidth]{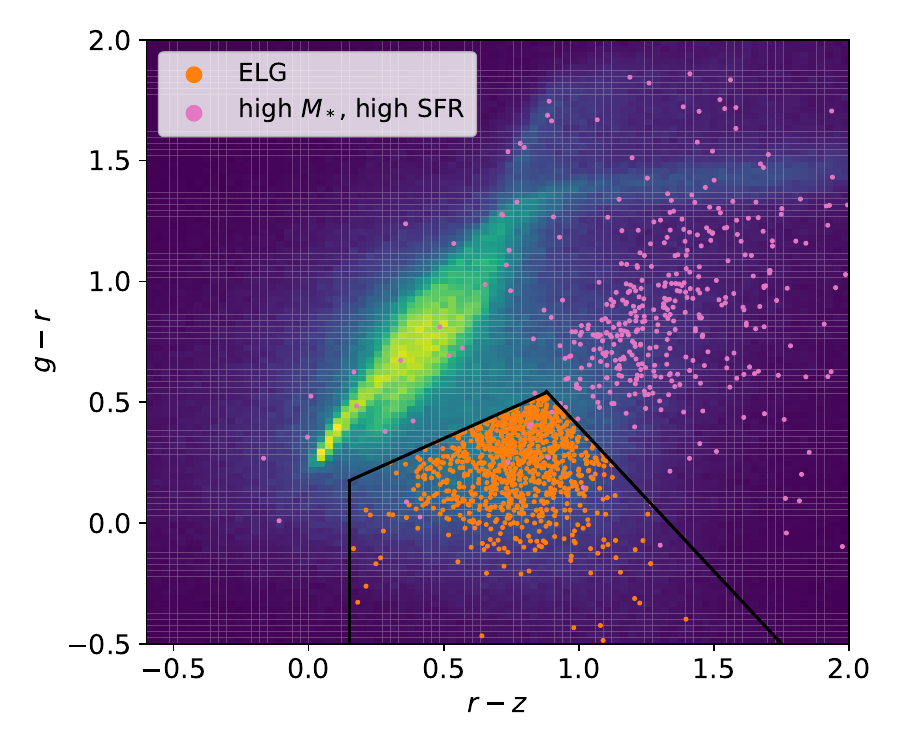}
    \vspace{-0.4cm}
    \caption{The $g-r$ vs $r-z$ distribution of the DESI ELGs in COSMOS compared to that of the most massive star-forming galaxies located in the region bounded by pink solid lines ($\log_{10}M_\mathrm{star} > 10.5$, $\log_{10}$SFR$>1.3$) in Figure~\ref{fig:completeness2d}. The 2D histogram shows the distribution of the full DESI Legacy DR9 sample. The black lines showcase the DESI EDR ELG selection cuts. Clearly, the most massive star-forming galaxies are rejected by the slanted cut with a negative slope. }
    \label{fig:selection}
\end{figure}

In Figure~\ref{fig:completeness2d}, the peak completeness of the DESI ELG sample is approximately $65\%$, occurring at approximately $\log_{10}M_\mathrm{star}/M_\odot = 9.8$ and the highest SFR end of the distribution. However, the most massive star-forming galaxies above $\log_{10}M_\mathrm{star}/M_\odot > 10.5$ are missing from the ELG sample. 
To understand this incompleteness at the massive end, we select these most massive star-forming galaxies with the pink bounding box in Figure~\ref{fig:completeness2d}. We cross-match them with DESI Legacy Survey DR9 catalogs \citep{2019Dey} to extract their photometry, and we plot their $g-r$ vs $r-z$ distribution in Figure~\ref{fig:selection}. The orange and pink scatter points show the ELG sample and the high stellar mass high SFR sample, respectively. The 2D histogram shows the distribution of the full Legacy Survey sample. The black lines show the DESI ELG selection \citep{2023Raichoor}. Clearly, the most massive star-forming galaxies are rejected by the negatively sloped selection cut. 
Referring to Figure~3 and Figure~18 of \cite{2023Raichoor}, this selection cut is chosen to optimize the fraction of ELG targets with high [O\,II] flux in $1.1 < z < 1.6$, which translates to high redshift success efficiency. The redder region occupied by the pink points is mostly populated with lower redshift galaxies. Thus, including this region of the color space would translate to a significantly lower redshift efficiency.

COSMOS photo-$z$s should be fairly accurate given that we found 929 out of the 931 ELG matches in COSMOS have photo-$z$ errors less than 0.1. Thus, most of the galaxies selected by the pink bounding box should be indeed high redshift massive galaxies. The reason they have redder colors can be attributed to their older ages, which are inferred to be 1.5$\pm 0.8$ billion years compared to the ELGs' $0.6\pm 0.4$ billion years. As a result, we suggest that these galaxies are redder because of their higher metallicities and also likely dust. We conduct similar tests for galaxies that are in the same region of the stellar mass--SFR plane but are not selected as ELGs. We find these galaxies are also redder in color, likely due to dust attenuation. 

It is also worth noting that the ELG sample we are considering is also at lower redshift $0.8 < z < 1.1$, outside the range that is optimized for ELG target selection. We reserve a detailed discussion of these points for a future paper. The conclusion, for now, is that while ELGs are indeed massive starburst galaxies, a significant fraction of the most massive star-forming galaxies are rejected by the ELG selection cuts in order to achieve higher redshift efficiency in $1.1 < z < 1.6$. 

\subsection{ELG morphology}
\label{sec:morph}
\begin{figure*}
    \hspace{-0.1cm}
    \includegraphics[width=1\textwidth]{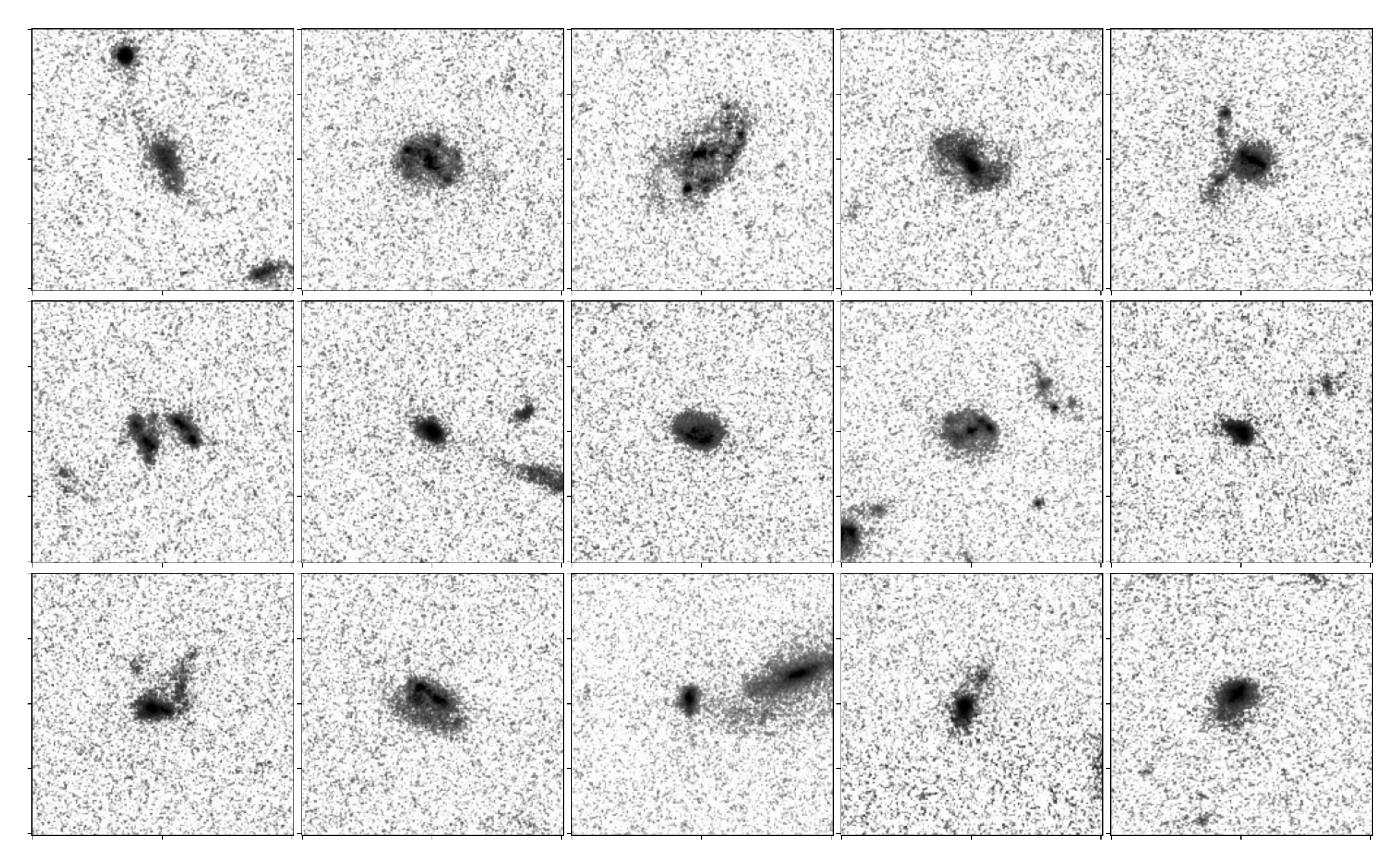}
    \vspace{-0.2cm}
    \caption{HST ACS (F814W) images of DESI ELGs in COSMOS. Each image stamp is 6 arcsec per side. Compared to Figure~\ref{fig:cosmos_stamps_notelg}, which shows galaxies of similar mass that are not identified as ELGs, ELGs show significantly more irregular morphology. ELGs also tend to appear as recently merged or part of soon-to-be-merged systems. This is consistent with our hypothesis that ELGs exhibit star formation triggered by tidal interactions.}
    \label{fig:cosmos_stamps_elg}
\end{figure*}
\begin{figure*}
    \hspace{-0.1cm}
    \includegraphics[width=1\textwidth]{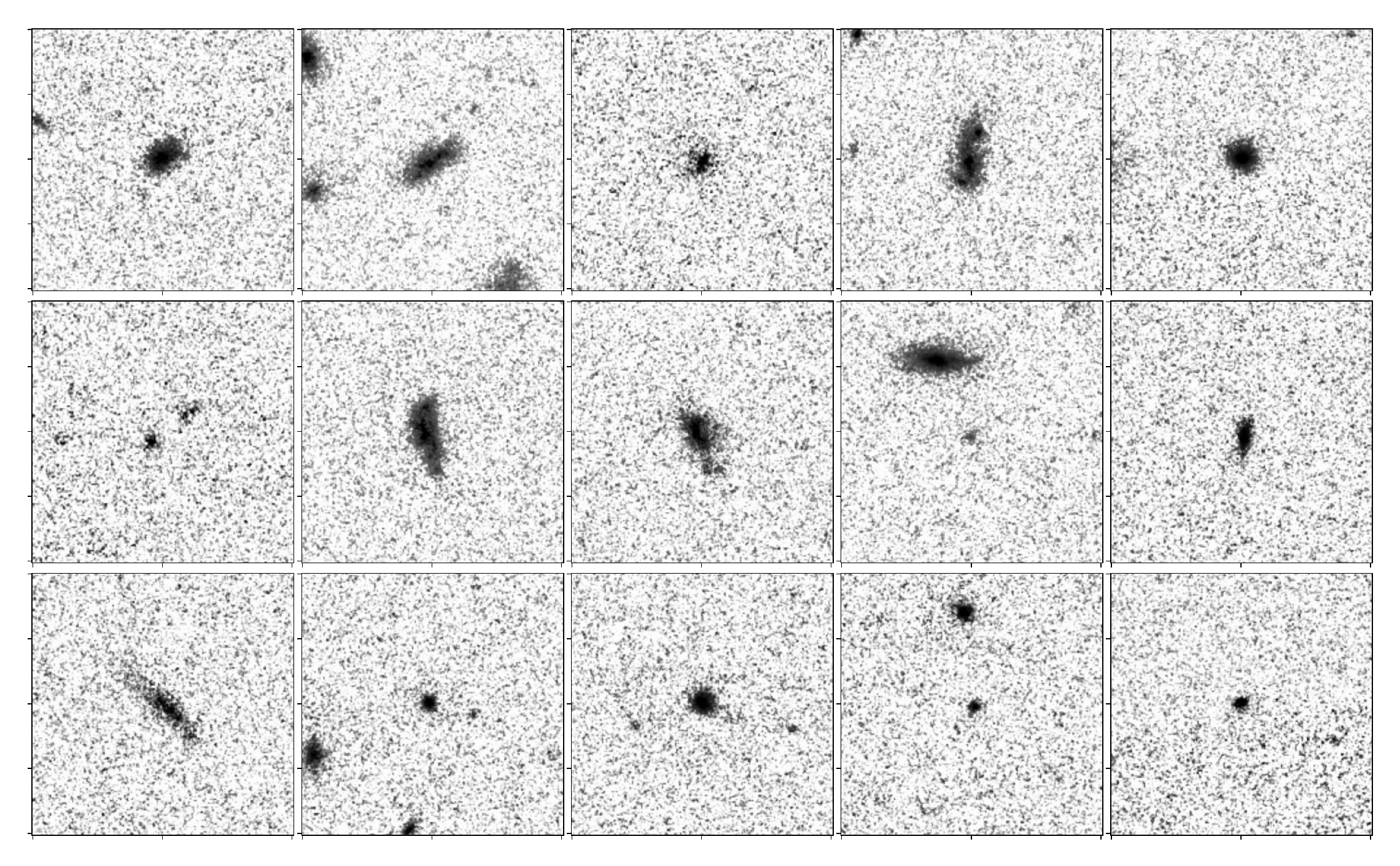}
    \vspace{-0.2cm}
    \caption{HST ACS (F814W) images of galaxies in COSMOS that are not identified as DESI ELGs but are of similar masses ($9.5 < \log_{10}(M_\mathrm{star}/M_\odot) < 9.7$). Each image stamp is a 6 arcsec per side. Compared to the ELGs in Figure~\ref{fig:cosmos_stamps_elg}, these galaxies tend to be more nuclear and less perturbed.}
    \label{fig:cosmos_stamps_notelg}
\end{figure*}

To study ELG morphology and build more physical intuition, we can also look at Hubble images of ELGs in COSMOS. 
Figure~\ref{fig:cosmos_stamps_elg} shows a random subset of HST ACS I-band (F814W) mosaic images of ELGs in COSMOS \citep{2007Koekemoer, 2007Scoville, 2009Massey}. These images have a pixel resolution of 0.03 arcsec and each image is 6 arcsec in size. 
For comparison, Figure~\ref{fig:cosmos_stamps_notelg} shows a random subset of galaxies in COSMOS that are of similar mass to DESI ELGs ($9.5 < \log_{10}(M_\mathrm{star}/M_\odot) < 9.7$) but are not identified as ELGs. 
Comparing the two sets of images, it is immediately clear that ELGs are more likely to have irregular/perturbed morphologies with strong tidal features. A similar preference for irregular morphology has also been seen for star-forming galaxies in the WiggleZ survey \citep{2010Drinkwater}. Moreover, while the non-ELGs appear fairly isolated, the ELGs tend to appear in the vicinity of other irregular galaxies. 

These signs of interaction may suggest that a significant fraction of ELGs are either recently merged or merging systems. This would also explain the high star formation rates of ELGs (Figures~\ref{fig:cosmos_completeness} and \ref{fig:completeness2d}), since both observations and simulations have long shown that galaxy mergers can trigger star formation \citep[e.g.][]{1978Larson, 1987Kennicutt, Knapen_2009, 2012Lambas, 2013Ellison, 2015Knapen, 2021Moreno, 2022Thorp}. It is believed that such star formation stems from gas losing angular momentum and fueling centralized star formation, as a result of the non-axisymmetric structures generated by gravitational forces in an interaction \citep[e.g.][]{1996Mihos, 2013Hopkins}. Though the enhancement of star formation can strongly depend on the specific configuration of the merger, orbital parameters, mass ratios, local densities, and gas contents of the disks \citep[e.g.][]{2008Cox, 2010Lin, 2015Davies, 2015Moreno, 2017Fensch}, at least some mergers can also trigger short periods of extreme star formation, known as starbursts, due to tidal interactions compressing and shocking the gas \citep[e.g.][]{2004Barnes, 2009Kim, 2009Saitoh, 2017Cortijo, 2023Verrico}. Thus, our picture of merger-triggered star formation in ELGs is solidly grounded in our existing understanding of galaxy evolution. 

\begin{figure}
    \hspace{-0.3cm}
    \includegraphics[width=0.45\textwidth]{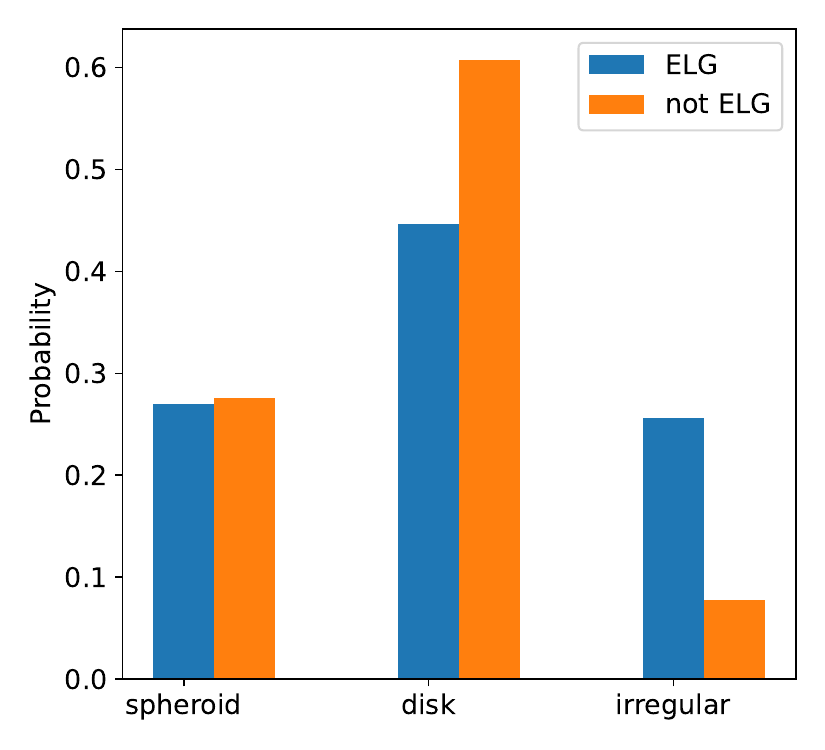}
    \vspace{-0.4cm}
    \caption{The mean morphology probability of ELGs and non-ELGs at similar masses classified by the \textsc{Morpheus} model using available HST images. The ELGs are more likely to be classified as irregular than non-ELGs at similar masses. }
    \label{fig:morph}
\end{figure}

We further substantiate our argument by cross-matching the ELG sample with other galaxy catalogs in the COSMOS field. 
We first consider the \textsc{Morpheus} catalog \citep{2019Hausen}, which uses a deep learning-based model to generate morphology classifications for galaxies on HST data in the CANDLES field. We found 20 matches within our ELG sample in $0.8 < z < 1.1$ and 89 non-ELG matches with stellar mass $9.5 < \log_{10}(M_\mathrm{star}/M_\odot) < 9.7$ for comparison. For each galaxy, the \textsc{Morpheus} catalog assigns probabilities (summing to unity) of having spheroid, disk, or irregular morphologies. We showcase the mean classification probabilities of ELGs and non-ELGs in Figure~\ref{fig:morph}. Compared to non-ELGs, the ELGs are more likely to be classified as irregular and less likely to be classified as disks, again suggesting that ELGs tend to be more disturbed than non-ELGs. 

To compare the merger rate in ELGs versus non-ELGs in a more quantitative way, we cross-match with the late-stage merger catalog compiled by \cite{2014Lackner}.
This catalog aimed to select galaxy mergers in their final stages before coalescence (i.e., when the galaxy nuclei are intact but at very small $\sim$kpc separations) by using a high-pass filter to identify multiple peaks in the surface brightness profiles of COSMOS HST ACS images.
We cross-match the ELG sample against this merger catalog and find that $(9.8\pm1.0)\%$ of ELGs are identified as late-stage mergers. Of a control sample of non-ELGs (in a similar mass and redshift range as the ELG sample), only $(0.97\pm0.32)\%$ are late-stage mergers. The errors reported here are standard deviations estimated by randomly drawing $10^4$ bootstrap subsamples from the ELG and non-ELG samples.

Importantly, the \cite{2014Lackner} catalog is incomplete in several ways: it is based on a source sample that is less complete than the COSMOS2020 catalog from which we draw our ELG sample, it only considers a narrow definition of ``late-stage'' mergers, and it may also miss up to $80\%$ of late-stage mergers at $z>0.5$ \citep[Appendix A of][]{2014Lackner}. As a result, the percentages of ELGs and non-ELGs in this catalog do not translate to absolute merger rates. However, the relative difference between the two classes---that ELGs are 10 times more likely to be identified as late-stage mergers than non-ELGs---further supports our argument that a significant fraction of ELGs are mergers.

The physical nature of the mergers in ELGs remains unclear. Returning to the cutouts in Figure~\ref{fig:cosmos_stamps_elg}, we see that the disky ELGs are accompanied by nearby galaxies that are in the process of merging or feeding the ELGs with tidal tails. We speculate that there are potentially multiple star formation triggers in ELGs depending on the environment and the mass ratio. The disky ELGs may be more dynamically stable, getting their star formation via gas accretion. In contrast, the irregular ELGs could be undergoing major or minor mergers that trigger starbursts via tidal compression. Further studies are needed to test this hypothesis. 

The merger-triggered starburst scenario naturally produces to 1-halo conformity in the ELG population. Given reservoirs of molecular gas, the tidal interactions induced by mergers can ignite star formation on both sides of the merger. As a result, ELG satellites would tend to appear in close pairs with ELG centrals, boosting the clustering on small scales. This type of conformity is also distinct from galaxy assembly bias as it is not directly tied to dark matter assembly, but is simply a result of proximity to neighbour galaxies and neighbour galaxy types. 



\section{HOD modeling}
\label{sec:hod}

A key objective of this paper is to explore ELG--halo connection models that describe the observed ELG clustering in detail. The Halo Occupation Distribution (HOD) is a simple yet powerful theoretical framework for predicting and interpreting galaxy clustering on non-linear scales. In this section, we review the basics of the HOD framework, motivate the need for extensions beyond the vanilla HOD model, and present definitions of conformity and galaxy assembly bias as plausible model extensions for ELGs. 

\subsection{Vanilla HOD}
The HOD model relates galaxies to their dark matter halo counterparts via a concise statistical relationship. In its simplest (vanilla) form, the HOD is summarized by $P(N|M)$, i.e. the probability of hosting $N$ galaxies given a halo of mass $M$. In the assumption of halo mass being the only relevant quantity, the HOD is a complete model of galaxy occupation. Historically, \cite{2004Kravtsov} was amongst the first to show that the HOD can accurately predict the clustering of galaxy-size subhalos down to $\sim 100h^{-1}$Mpc in a dark matter only N-body simulation. Then a series of studies demonstrated that a simple HOD model can describe the clustering of luminosity-limited galaxy samples at a variety of redshifts \citep[e.g.][]{2005Zehavi, 2006Conroy, 2011Zehavi}. At the same time, the same HOD framework was shown to also describe the clustering of color-magnitude selected LRG samples \citep[e.g.][]{2009Zheng, 2010Watson, 2013Parejko, 2016Rodriguez}. \cite{2007bZheng} and \cite{2008Coil} further demonstrated the validity of HOD modeling up to $z \sim 1$. 

LRGs are selected to have high completeness at the massive end, and their clustering has been shown to be well described by the following model \citep[e.g.][]{2005Zheng, 2007bZheng, 2015Kwan, 2021bYuan}:
\begin{align}
    \bar{n}_{\mathrm{cent}}^{\mathrm{LRG}}(M) & = \frac{f_\mathrm{ic}}{2}\mathrm{erfc} \left[\frac{\log_{10}(M_{\mathrm{cut}}/M)}{\sqrt{2}\sigma}\right], \label{equ:zheng_hod_cent}\\
    \bar{n}_{\mathrm{sat}}^{\mathrm{LRG}}(M) & = \left[\frac{M-\kappa M_{\mathrm{cut}}}{M_1}\right]^{\alpha}\bar{n}_{\mathrm{cent}}^{\mathrm{LRG}}(M).
    \label{equ:zheng_hod_sat}
\end{align}
In this model, $M_{\mathrm{cut}}$ characterizes the minimum halo mass to host a central galaxy. $M_1$ characterizes the typical halo mass that hosts one satellite galaxy. $\sigma$ describes the steepness of the transition from 0 to 1 in the number of central galaxies. $\alpha$ is the power law index on the number of satellite galaxies. $\kappa M_\mathrm{cut}$ gives the minimum halo mass to host a satellite galaxy.
We have added a modulation term $\bar{n}_{\mathrm{cent}}^{\mathrm{LRG}}(M)$ to the satellite occupation function to avoid scenarios where LRG satellites occupy halos not massive enough to host LRG centrals. We have also included an incompleteness parameter $f_\mathrm{ic}$, which is a downsampling factor controlling the overall number density of the mock galaxies. This parameter is relevant when trying to match the observed mean density of the galaxies in addition to clustering measurements. By definition, $0 < \mathrm{ic}\leq 1$.

ELGs are selected based on [O\,II] luminosity, which should trace star formation, and are known to have a rather low characteristic halo mass. Distinct HOD analytic forms have been developed for ELGs, and it was found that central models that show a peaked distribution with small to no contribution from the massive end tend to describe the observed clustering well \citep{2020Alam, 2020Avila}. 
For this paper, we adopt the mHMQ model from \cite{2023Rocher} as the vanilla model. This is a slightly modified version of the HMQ model presented in \cite{2020Alam}. Here we use a notation that is consistent with \cite{2022Yuan} but slightly different than \cite{2023Rocher}. The two notations are completely equivalent. 

In the vanilla model, the number of ELG centrals is given by 
\begin{align}
\bar{n}_{\mathrm{cent}}^{\mathrm{ELG}}(M) &=p_\mathrm{max}\phi(M) \Phi(\gamma M),
\end{align}
where, 
\begin{align}
\phi(x) &=\mathcal{N}\left(\log _{10} M_{\mathrm{cut}}, \sigma_M\right) \\
\Phi(x) &=2\int_{-\infty}^x \phi(t) d t=1+\operatorname{erf}\left(\frac{x}{\sqrt{2}}\right).
\label{equ:elg_cent}
\end{align}
We refer to \cite{2020Alam} for detailed descriptions of each parameter. Briefly, $p_\mathrm{max}$ controls the peak completeness in the central occupation. $\phi(M)$ defines Gaussian distribution with peak position determined by $M_\mathrm{cut}$ and width determined by $\sigma_M$. $\Phi(\gamma M)$ introduces asymmetry to the Gaussian with a tilt parameter $\gamma$. 

For the satellite galaxies, we adopt a power law model,
\begin{equation}
\bar{n}_{\mathrm{sat}}^{\mathrm{ELG}}(M) = \left[\frac{M-\kappa M_{\mathrm{cut}}}{M_1}\right]^{\alpha},
\label{equ:elg_sat}
\end{equation}
where the parameters carry similar meanings as the LRG satellite case. Note that we do not modulate with $n_\mathrm{cent}$ because ELG selection is driven by bursty star formation, so we want to allow for the possibility of ELG satellites existing in halos less massive than those hosting ELG centrals. 

In the vanilla HOD models, the position and velocity of the central galaxy are set to be the same as those the halo center, specifically the L2 subhalo center-of-mass for the {\sc CompaSO} halos. For the satellite galaxies, they are randomly assigned to halo particles with uniform weights, each satellite inheriting the position and velocity of its host particle. We defer the readers to \cite{2021bYuan} for implementation details. 

\subsection{Conformity and galaxy assembly bias}
\label{subsec:conf_gab}
\cite{2023Rocher} showed that the vanilla HOD struggles to describe the ELG auto-correlation function observed in DESI at small scales, and that extensions to the vanilla HOD models are needed. Recent studies using simulated galaxy models have identified galaxy assembly bias and galaxy conformity as secondary effects that are both well motivated and can potentially explain the observed clustering \citep[e.g.][]{2023Contreras, 2022mHadzhiyska, 2022Wang, 2022Beltz-Mohrmann, 2021Yuan}. 
In this sub-section, we introduce these two effects and offer succinct definitions in the context of this paper. 

\textbf{Galaxy assembly bias} is defined as follows: At fixed halo mass, the galaxy properties or number of galaxies within dark matter halos may depend on secondary halo properties. In an HOD, this is expressed as $P(N|M, x)$, where $x$ is some secondary halo property beyond halo mass that is correlated with the assembly history of the halo \citep{2018Wechsler}. 
Galaxy assembly bias is relevant for galaxy clustering analyses because halo clustering depends on halo assembly history (an effect known as halo assembly bias). Thus, galaxy assembly bias would lead to a signature in galaxy clustering as well \citep[e.g.][]{2005Gao, 2005Zentner, 2006Wechsler, 2007Gao, 2007Croton, 2008Li, 2016Miyatake}.
There is an extensive literature studying galaxy assembly bias in simulations and data in the last decades \citep[e.g.][]{2006Zhu, 2014Zentner, 2014Pujol, 2016Hearin, 2018Artale, 2018Zehavi, 2019Bose, 2021Yuan, 2019Lange, 2019Contreras, 2020Xu, 2022Wang, 2022Salcedo, 2022Beltz-Mohrmann}. For a review of this topic, we refer the reader to \cite{2018Wechsler}. 

In this paper, \textbf{Conformity} refers to the phenomenon where the properties or number of satellites depends on the property of the central galaxy. In an HOD, conformity can be expressed as $P(N_\mathrm{satellite}|M, x_\mathrm{central})$, where $x_\mathrm{central}$ is some property of the central galaxy, such as galaxy type or color. The conformity definition is considerably more strict than historical notions of galactic conformity, which generically refers to some level of spatial correlation in galaxy properties \citep[e.g.][]{2006Weinmann, 2013Kauffmann, 2015Hearin}. Our definition specifically refers to a 1-halo central-satellite conformity, as opposed to 2-halo conformity where the properties of galaxies in adjacent halos might be correlated, or a 1-halo satellite-satellite conformity where the properties of satellites in the same halo might be mutually dependent. Discussion of these other types of conformity are interesting but beyond the scope of this paper.

Conformity in certain forms have been regarded as an emergent phenomenon of halo and galaxy assembly bias \citep[e.g.][]{2015Hearin, 2016bHearin, 2017Berti}. Specifically, \cite{2016bHearin} argued that 2-halo conformity is a direct result of correlated halo accretion rate, which arises because neighboring halos are subject to the same large-scale tidal field. Strong tidal effects inhibit mass accretion into halos, resulting in spatially correlation star formation. The same paper argued that 1-halo conformity can be a result of 2-halo conformity at higher redshift. If halo accretion is the sole origin of conformity at fixed halo mass, then conformity is simply an alternative quantification of galaxy assembly bias. 

However, a key point of this paper is to show that the 1-halo conformity we find is not necessarily degenerate with galaxy/halo assembly bias. In our definition, $P(N_\mathrm{satellite}|M, x_\mathrm{central})$, conformity is only degenerate with galaxy assembly bias if the relevant central property $x_\mathrm{central}$ is fully determined by the dark matter assembly of the host halo. In fact, as we show in section~\ref{sec:cosmos}, section~\ref{sec:tng}, and section~\ref{sec:discuss}, the conformity we see in DESI ELGs is likely a result of merger-driven starburst, which is not a direct result of halo accretion. This type of conformity is separate from galaxy assembly bias. 

We present our implementations of galaxy assembly bias and conformity in section~\ref{sec:desi_method}.

\section{Conformity and assembly bias in \textsc{IllustrisTNG}}
\label{sec:tng}

We first seek evidence for conformity and galaxy assembly bias and test their effects on clustering in the publicly available \tng\ hydrodynamical simulation. Hydrodynamical simulations are valuable because they self-consistently model galaxy evolution through detailed recipes of stellar evolution, various feedback process, magnetic fields, and etc. They allow us to statistically study galaxy evolution in a physically plausible setting. They are particularly well suited for probing gaps in our galaxy evolution theory and building up more sophisticated models. 

\subsection{\tng}
\textsc{IllustrisTNG} is a state-of-the-art magneto-hydrodynamic simulation suite \citep[][]{2018Pillepich,2018Marinacci,2018Naiman,
2018Springel,2019Nelson,2018Nelson,2019Pillepich,2019bNelson}. 
 \textsc{IllustrisTNG} were carried out using the \textsc{AREPO}
code \citep{2010Springel} with cosmological parameters consistent with the
\textit{Planck 2015} analysis \citep{2016Planck}.
These simulations feature a series of improvements
compared with their predecessor, \textsc{Illustris}, such as
improved kinetic AGN feedback and galactic wind
models, as well as the inclusion of magnetic fields.

In particular, we utilize the \textsc{IllustrisTNG}-300-1 box, the largest high-resolution hydrodynamical simulation from the suite. 
The size of its periodic box is 205$h^{-1}$Mpc with 2500$^3$ DM particles
and 2500$^3$ gas cells, implying a DM particle mass of $3.98 \times 10^7 \ h^{-1}\rm{M_\odot}$ and
baryonic mass of $7.44 \times 10^6 \ h^{-1}\rm{M_\odot}$. 
We also use the dark-matter-only (DMO) counterpart of the \textsc{IllustrisTNG}-300-1 box, \textsc{IllustrisTNG}-300-Dark, which was evolved with the same initial conditions and the same number of dark matter particles ($2500^3$), each with particle mass of $4.73\times 10^7 h^{-1}M_\odot$.

The haloes (groups) in \textsc{IllustrisTNG}-300-Dark are found with a standard 
friends-of-friends (FoF) algorithm with linking length $b=0.2$ (in units of the mean inter-particle spacing)
run on the dark matter particles, while the subhaloes are identified 
using the SUBFIND algorithm \citep{2000Springel}, which detects 
substructure within the groups and defines locally overdense, self-bound particle groups.
For this paper, we analyse the simulations at redshift $z = 0.8$. 

\subsection{Mock sample selection}

To select DESI-like galaxies, we first calculate DESI-like photometry for all \tng\ galaxies. The technical details are described in \cite{2022Yuan} and \cite{2021eHadzhiyska}. Essentially, we use the Flexible Stellar Population Synthesis code to calculate raw galaxy fluxes in each DESI photometric bands \citep[FSPS,][]{2010Conroy}. Then we obtain the final DESI magnitudes by feeding the raw fluxes through an empirical model of dust attenuation, where the model parameters are roughly calibrated on recent observational measurements. Finally, we apply DESI color-magnitude cuts to obtain mock LRG and ELG samples \citep{2020Zhou, 2022Zhou, 2020Raichoor, 2023Raichoor}. We select 4608 LRGs, corresponding to a density of approximately $5\times 10^{-4}h^3$Mpc$^{-3}$; and 8637 ELGs, corresponding to a density of roughly $1\times 10^{-3}h^3$Mpc$^{-3}$. 



\subsection{Galaxy assembly bias tests}
\label{subsec:tng_ab}



There have been a series of studies that have charaterized the effect of galaxy assembly bias in \tng.  We summarize those results here.

For LRGs, the predominant signal is a correlation with a secondary property that connects to local overdensity on scales of a few megaparsecs. \cite{2022Yuan} and \cite{2020Hadzhiyska} found that HODs parameterized on halo mass and local environment are able to recover the LRG clustering seen in \tng\ LRG samples, and ignoring the secondary dependency can result in a significant bias in clustering prediction between $10\%$ and $20\%$. \cite{2021Delgado} compared different secondary halo properties and found the local environment to be the best tracer of secondary biases. More recently, \cite{2022mHadzhiyska} confirmed these results using the larger \textsc{MilleniumTNG} simulations \citep[MTNG;][]{2022Hernandex-Aguayo, 2022Bose, 2022Pakmor}, while \cite{2021Xu} found similar trends in semi-analytic galaxy mocks. \cite{2021Yuan} found the BOSS CMASS LRG sample to also favor the inclusion of an environment-based galaxy assembly bias model.

For ELGs, the galaxy assembly bias signal is much less certain, especially given the limited volume of \tng. \cite{2021eHadzhiyska} found a $\sim 4\%$ bias in clustering prediction when ignoring any secondary bias, compared to the 10-20$\%$ for LRGs. With MTNG, \cite{2022mHadzhiyska} found a similar but statistically significant signal of 5-10$\%$ bias in the $z = 1$ ELG clustering. \cite{2022mHadzhiyska} compared different secondary halo properties and found that the local shear calculated on scales of approximately $1h^{-1}$Mpc to be the best tracer of ELG assembly bias. Beyond hydrodynamical simulations, \cite{2021Jimenez} found scale-dependent ELG assembly bias signature in a semi-analytic model approach.


\subsection{Conformity tests}

In this section, we probe for central-satellite conformity in mock LRG and ELG samples in \tng. First of all, conformity is irrelevant for LRG satellites as all LRG satellites living in high mass halos that are essentially guaranteed to host LRG centrals. Thus, conformity is only relevant for modeling the occupation of ELG satellites. 

\begin{figure}
    \hspace{-0.2cm}
    \includegraphics[width=0.48\textwidth]{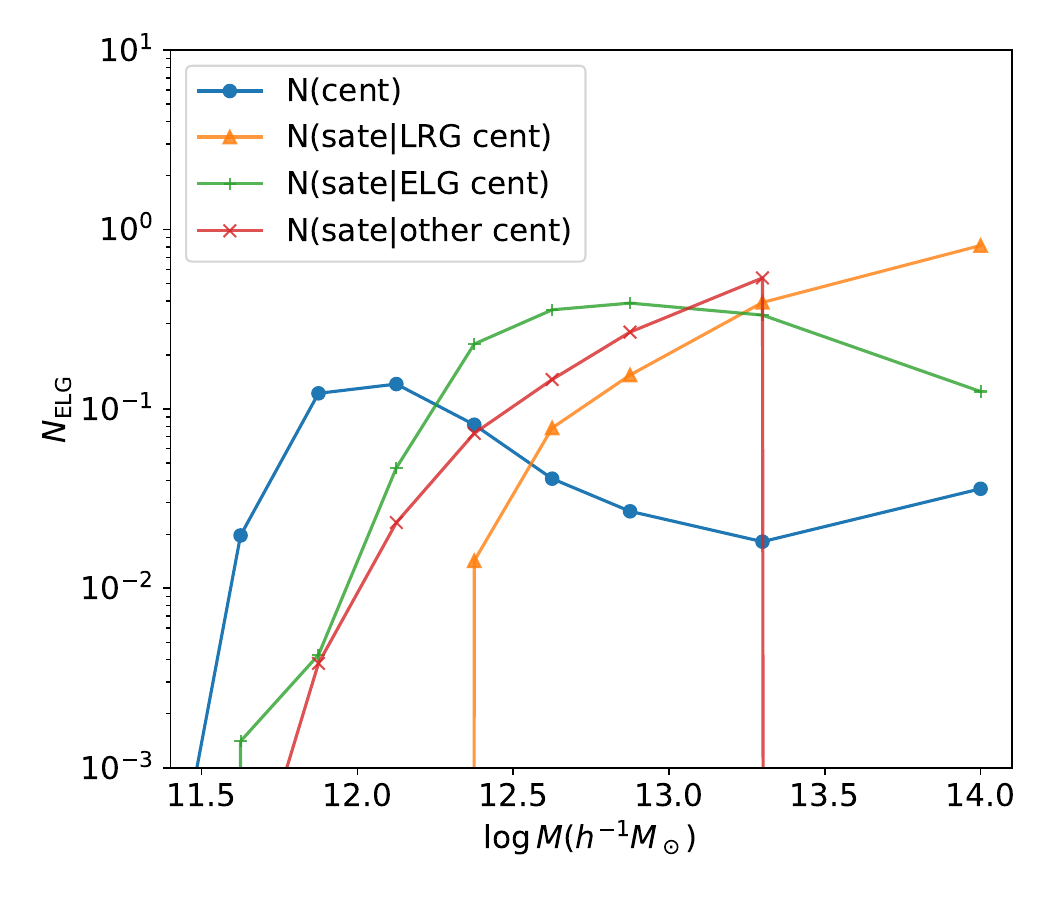}
    \vspace{-0.2cm}
    \caption{The mean HOD of the \textsc{IllustrisTNG} ELG sample. We break the satellite occupation distribution into three conditional probabilities. The green curve shows the mean number of ELG satellites when the halo also hosts an ELG central. The orange curve shows the number of ELG satellites when the halo hosts an LRG central. The red curve shows the case when the central is neither an ELG or LRG. We see clear signatures of conformity. }
    \label{fig:hod_elg}
\end{figure}

Figure~\ref{fig:hod_elg} shows the ELG HOD as measured in the \tng\ sample. The blue line shows the occupation of the ELG centrals. The satellite occupation is broken into three scenarios: satellites around and ELG central in green; satellites around LRG central in orange; and satellites around centrals that are neither in red. There is clear indication of conformity: at fixed halo mass, the number of satellites is largest when the central is an ELG, and lowest when the central is an LRG. At the largest halo mass, the green line unexpectedly drops below the orange line, but we suspect it is simply because of low signal-to-noise due to the limited sample size at high masses. We also expect ELG centrals to be quite rare at the high mass end and that they should not contribute significantly to the clustering. 

\cite{2022Yuan} showed that the vanilla HOD for \tng\ ELGs can be well described by a skewed Gaussian model for centrals and a power law model for the satellites (See Equation 9-13 of said paper and also \cite{2020Alam}). Judging by Figure~\ref{fig:hod_elg}, a simple approach to modeling satellite-central conformity is simply by scaling the satellite occupation as a function of the central galaxy type. 
We will construct such models and test them against DESI data in section~\ref{sec:desi_method}-\ref{sec:desi}.

\begin{figure*}
    \hspace{-0.7cm}
    \includegraphics[width=1\textwidth]{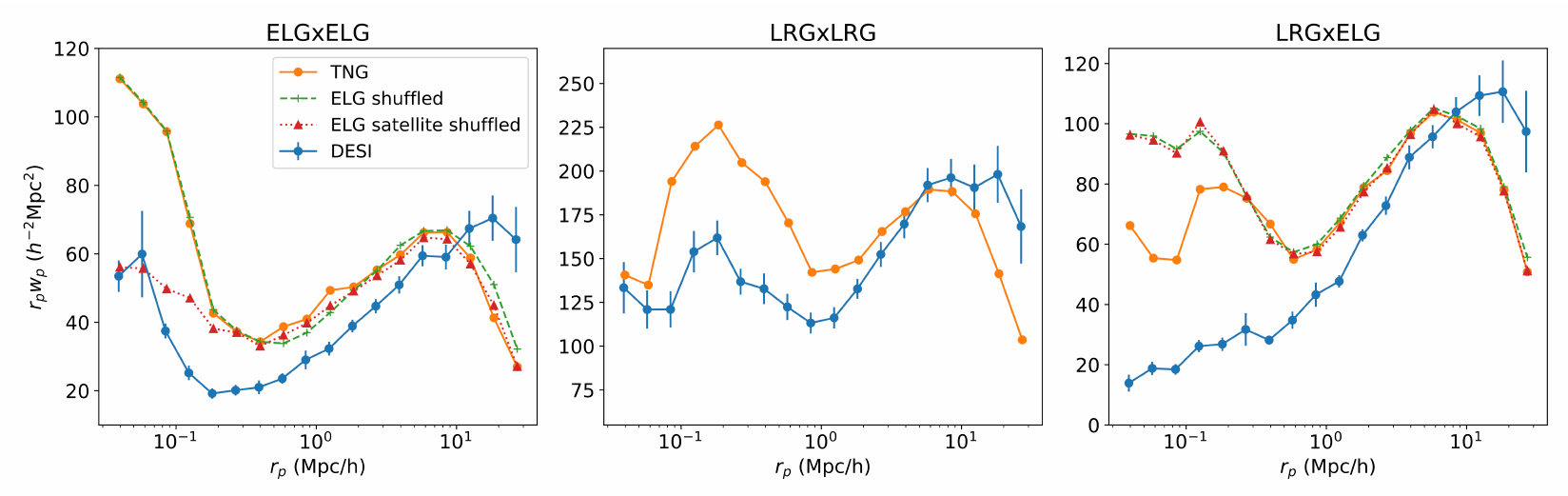}
    \vspace{-0.2cm}
    \caption{The clustering of mock LRG and ELG samples selected in \textsc{IllustrisTNG} (orange), and the effects when two different shuffling routines are applied to the ELG sample in \textsc{IllustrisTNG} (green and red). Green curves shuffle both the ELG central and satellites to test for 2-halo conformity and galaxy assembly bias, whereas the red curves shuffle just the ELG satellites while holding the centrals fixed. The left panel shows a significant decrease in small-scale amplitude when only the ELG satellites are shuffled, suggesting ELG satellites preferentially occupy halos with ELG centrals. The right panel shows the small-scale amplitude increasing when the ELGs are shuffled, suggesting that ELG satellites preferentially occupy halos without LRG centrals. The changes in clustering on very small scales suggest strong 1-halo conformity in the ELG sample. The fact that shuffling does not alter clustering on $\sim 10h^{-1}$Mpc scales suggests the lack of 2-halo conformity. }
    \label{fig:wps_shuffle}
\end{figure*}

To test how conformity affects clustering, we conduct several shuffling tests similar to those used in \cite{2022Yuan, 2020Hadzhiyska}. Specifically, we shuffle the galaxy contents amongst halos of the same mass and compare the clustering of galaxy sample before and after the shuffling. This shuffling procedure removes any secondary dependency of the galaxy--halo connection one anything other than halo mass---such as a dependency on galaxy type of nearby galaxies, as predicted by conformity. Thus any change in the clustering due to shuffling would indicate the effect of one or more secondary dependencies on galaxy clustering.

Figure~\ref{fig:wps_shuffle} showcases the projected auto and cross-correlations of the \tng\ LRG and ELG mock samples before shuffling in orange and post shuffling in green and red. We also show the DESI auto and cross correlation measurements in blue for comparison. We approximate the error bars on the \tng\ samples as simply the DESI jackknife errors re-scaled by the \tng\ sample size. We see that the amplitude of the mock clustering roughly agrees with the DESI One-Percent data, and the shape of the mock auto-correlations produce the right qualitative features. This suggests that our LRG and ELG mock samples are reasonable approximations of the corresponding DESI samples. \tng\ ELGs also produce a similar upturn at small separation in the auto-correlation function. However, there is a large deviation in the cross-correlation measurement, where the DESI measurement shows a much stronger anti-correlation between the ELG and LRG samples. This could be due to insufficient quenching of blue galaxies near red galaxies in \tng, or due to systematics in the modeling of dust attenuation and sample selection.

The green dashed curves in Figure~\ref{fig:wps_shuffle} show the clustering of the sample after shuffling ELGs amongst halos of the same mass. Specifically, we take all ELGs (both centrals and satellites) in a halo, and randomly assign them to another halo of the same mass, while preserving the relative positions of the ELGs within the same halo, and we repeat this process for all halos. This test serves to check for any indications of galaxy assembly bias or 2-halo conformity in the ELG sample. 
The red dotted curves show the clustering of the sample after shuffling only the ELG satellites amongst halos of the same mass. Shuffling only the satellites while holding the centrals in place disrupts any 1-halo central-satellite conformity that might exist. Thus, any changes in the clustering as a result suggests that the central and satellite occupations are not independent and the existence of 1-halo conformity.

On the left hand panel, the dashed green curve agrees perfectly with the unshuffled sample shown in orange. This suggests that the mock ELG sample does not show significant net effect of galaxy assembly bias and 2-halo conformity at the precision available with \tng. While this is contradictory to the evidence for ELG assembly bias found in MTNG and semi-analytic models, we are likely limited by the small volume of \tng. We continue to search for evidence of ELG assembly bias in DESI data in section~\ref{sec:desi}. However, the dotted red curve shows that the satellite shuffling significantly reduces the amplitude of ELG auto-correlation function on 1-halo scales. This suggests a strong positive conformity between ELG centrals and satellites in \tng. The fact that this effect only shows up when shuffling satellites but not when shuffling both centrals and satellites together demonstrates that the 1-halo conformity is independent from assembly bias. The observed DESI ELG auto-correlation also shows a large 1-halo upturn similar to the \tng\ ELGs, suggesting that a strong ELG conformity signature might also exist in the DESI ELG sample. We will test this claim in section~\ref{sec:redmapper}-\ref{sec:desi}.

On the LRG$\times$ELG panel on the right panel of Figure~\ref{fig:wps_shuffle}, the green dashed line is also consistent with the orange line on scales beyond $r_p\sim 1h^{-1}$Mpc. However, on smaller scales, the shuffled sample shows a larger clustering amplitude than the unshuffled sample. The satellite shuffling also shows an identical effect as shuffling the full ELG sample. This suggests a negative 1-halo conformity between LRGs and ELGs, specifically, ELG satellites disfavor halos with LRG centrals. This fits our picture of ELGs as merger-driven starbursts, as LRG centrals are biased tracers and tend to live in denser environments than ELGs. In denser environments, merger rates are low due to tidal stripping, and the mergers that do happen in dense environments tend to be ``dry'' (meaning no molecular gas) and thus do not trigger (or even suppress) star formation \citep[][]{2008Tran, 2010Lin, 2010Hester}. 

\begin{figure}
    \hspace{-0.2cm}
    \includegraphics[width=0.48\textwidth]{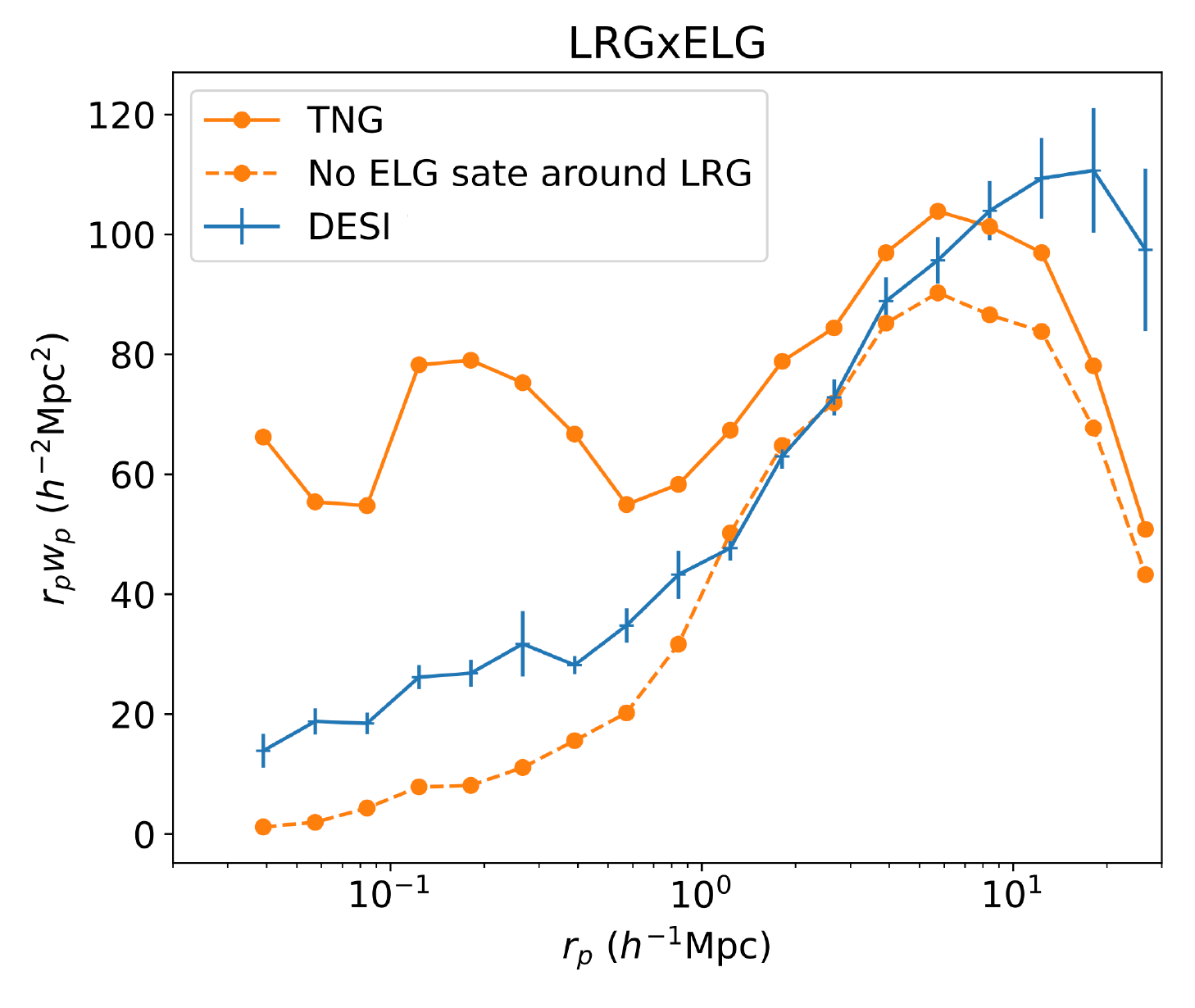}
    \vspace{-0.2cm}
    \caption{The effect of removing all ELG satellites with LRG centrals on the projected LRG$\times$ELG cross-correlation function. The solid orange line shows the clustering of the \tng\ ELG sample, and the dashed orange line shows the clustering of the same sample but removing all ELG satellites that are in halos with LRG centrals. The dashed orange line shares a similar shape as the observed DESI cross-correlation, albeit with even low clustering amplitude on 1-halo scales. This suggests that the ELG and LRG sample in DESI are likely more anti-correlated than seen in \tng.}
    \label{fig:wpel_nosate}
\end{figure}

Comparing the LRG$\times$ELG cross-correlations in \tng\ with DESI in blue shows that the two samples might be significantly more anti-correlated in data, suggesting that massive red centrals may turn off nearby star formation much more effectively than we thought. To illustrate this point more, the dashed curve in Figure~\ref{fig:wpel_nosate} shows the cross-correlation measurement in \tng\ when we remove all ELG satellites in halos with LRG centrals. The fact that the DESI measurement lies in between the two orange lines on small scales suggests that while there are still ELG satellites around LRG centrals in data, the data display stronger anti-conformity between ELG satellites and LRG centrals than predicted by \tng. 

\section{DESI HOD analysis framework}
\label{sec:desi_method}

Given the evidence for conformity and galaxy assembly bias in \tng, we now model these effects in a multi-tracer HOD framework and test the model against DESI auto- and cross-correlation data. We first describe the model and methodology of the HOD analysis in this section, and then we present the results in the next section.

\subsection{\textsc{AbacusSummit} simulations}

To model the underlying dark matter density field, we use the \textsc{AbacusSummit} simulation suite, which is a set of large, high-accuracy cosmological N-body simulations using the \textsc{Abacus} N-body code \citep{2021Maksimova, 2019Garrison, 2021bGarrison}. This suite is designed to meet the Cosmological Simulation Requirements of the Dark Energy Spectroscopic Instrument (DESI) survey \citep{2013arXiv1308.0847L}. For this analysis, we use the $z=0.8$ snapshot of the \verb+AbacusSummit_base_c000_ph000+ box, which contains $6912^3$ particles within a $(2h^{-1}$Gpc$)^3$ volume, yielding a particle mass of $2.1 \times 10^9 h^{-1}M_\odot$. The simulations relevant for this analysis assumes Planck 2018 cosmology. \footnote{For more details, see \url{https://abacussummit.readthedocs.io/en/latest/abacussummit.html}} 

The \textsc{AbacusSummit} also uses the {\sc CompaSO} halo finder, which is a highly efficient on-the-fly group finder  \citep{2021Hadzhiyska}. 
{\sc CompaSO} builds on the existing 
spherical overdensity (SO) algorithm
by taking into consideration the tidal radius
around a smaller halo before competitively
assigning halo membership to the particles
in an effort to more effectively deblend halos.

\subsection{\ahod}

We use the highly optimised \ahod\ package to forward model galaxy mocks and compute galaxy clustering \citep[][]{2021bYuan}. The code is highly efficient and is specifically designed for repeated HOD evaluations on \textsc{AbacusSummit} boxes. The package enables multi-tracer galaxy assignment and incorporates a range of physically motivated HOD extensions, such as conformity and assembly bias. The code is publicly available as a part of the \textsc{abacusutils} package at \url{https://github.com/abacusorg/abacusutils}. Example usage can be found at \url{https://abacusutils.readthedocs.io/en/latest/hod.html}.

We described the vanilla mass-only HOD models in section~\ref{sec:hod}. Here we describe introduce extensions that model effects of galaxy assembly bias and conformity. To include assembly bias into the HOD model, we follow the \ahod\ implementation where $\log_{10} M_\mathrm{cut}$ and $\log_{10} M_1$ are modified through the following parameterization:
\begin{align}
 \log_{10} M_{\mathrm{cut}}^{\mathrm{mod}} & = \log_{10} M_{\mathrm{cut}} + B_\mathrm{cent}(x^{\mathrm{rank}} - 0.5), \\
 \log_{10} M_{1}^{\mathrm{mod}} & = \log_{10} M_{1} + B_\mathrm{sat}(x^{\mathrm{rank}} - 0.5),
 \label{equ:AB}
\end{align}
where $x$ denotes the secondary property used to trace assembly bias. For LRGs, we use the local environment defined as the mass enclosed within 5$h^{-1}$Mpc of the halo (excluding the halo itself). For ELGs, we use the local shear as defined in Equation~7 of \cite{2022mHadzhiyska}. The superscript ``rank'' refers to the normalization procedure where the environment or the shear is ranked amongst halos in each mass bin and then the rank is normalized to a number between 0 and 1. For the shear calculation, we use $R = 0.5\,h^{-1}$Mpc and a grid size of $1000^3$.

To model central-satellite conformity, we propose a simple model directly inspired by Figure~\ref{fig:hod_elg}. We introduce two extension parameters to the vanilla HOD to model ELG-ELG conformity and ELG-LRG conformity, respectively. Specifically, we modulate the $M_1$ parameter, which controls the overall amplitude of satellite occupation:

\begin{equation}
    M_1^\mathrm{mod} = 
    \begin{cases}
    M_\mathrm{1, EE} & \textrm{if ELG central}\\
    M_\mathrm{1, EL} & \textrm{if LRG central}\\
    M_\mathrm{1} & \textrm{if neither},\\
    \end{cases}
    \label{equ:elgconfhod}
\end{equation}
where $M_{1,EE}$ modulates the ELG-ELG conformity strength and $M_{1,EL}$ modulates the ELG-LRG conformity strength. 
This model is more flexible than the maximal conformity model implemented in \cite{2023Rocher}, where all ELG satellites are assigned to halos with ELG centrals, which is equivalent to setting $M_1 = \infty$ in our model. This description also effectively distinguishes our conformity model from the galaxy assembly bias model. While the assembly bias model correlates satellite occupation to halo properties and local density, the conformity model only depends on the central galaxy type. Our modeling is also consistent with our proposed physical picture. One can interpret assembly bias as how galaxy color depends on halo assembly history, whereas conformity is largely driven to triggered star formation. \footnote{All \ahod\ model ingredients referenced in this paper are implemented and released on the official GitHub \url{https://github.com/abacusorg/abacusutils}.}

\subsection{Likelihood modeling}

For this analysis, we choose the projected 2PCF as our summary statistics. The projected 2PCF marginalizes over the velocity space features of the redshift-space 2PCF, thus removing the need to model galaxy velocities and avoiding potential systematic biases and degeneracies. We compute the projected 2PCF from the galaxy mocks using the fast \textsc{Corrfunc} code \citep{2020Sinha}.



Then we construct a Gaussian log-likelihood function as follows,
\begin{equation}
    \log L =  -\frac{1}{2}\sum_{i = 0}^n\left(\frac{\xi_{i, \mathrm{mock}} - \xi_{i, \mathrm{data}}}{\sigma_i}\right)^2 - \frac{1}{2}\frac{(\bar{n}_\mathrm{mock}-\bar{n}_\mathrm{data})^2}{\sigma_n^2}.
    \label{equ:logl}
\end{equation}
where $\xi_i$ is the $i$-th of the data vector in consideration. $\sigma_i$ denotes the jackknife error computed on the data. We ignore the off-diagonal covariances in this analysis because of the small number of jackknife regions and the relatively large data vector. It is also the case that the statistical variance at highly non-linear scales is dominated by shot noise coming from stochastic galaxy occupation, which is de-correlated over scale. We intend to confirm the key findings of this analysis with full covariances when a large data set becomes available. We have also added a term for number density, where $\sigma_n$ is the uncertainty on the measured mean number density. We assume $\sigma_n/\bar{n}_\mathrm{data} = 10\%$ \citep[][]{2022Yuan, 2015aGuo}.

Having defined our simulations, HOD models, and likelihood functions, we can sample the parameter space to find the best-fit models to the data. We use a global optimization routine called Covariance matrix adaptation evolution strategy (CMA-ES) with 400 random walkers until the walkers converge \citep{stochopy}. Then we repeat this process 5 times with different random seeds to ensure global convergence. 

\section{DESI HOD analysis results}
\label{sec:desi}
In this section, we present the main findings of DESI LRG$\times$ELG analyses. First, we fit our extended HOD models on only the LRG$\times$LRG and ELG$\times$ELG auto-correlation functions while using the cross-correlation function as a predictive test. Then, we rerun our fits to also include the cross-correlation function to test the final goodness-of-fit of the favored model. 

Because of the large parameter space, a very large number of likelihood calls are needed to achieve convergence in the optimization. To reduce the number of evaluations needed, we perform a two step optimization. First, we optimize the LRG HOD parameters against the LRG auto-correlation function. Then, we fix the LRG HOD parameters at their best-fit values, and we optimize the ELG HOD parameters against the ELG auto-correlation function in section~\ref{subsec:auto}, or together with the LRG$\times$ELG cross-correlation in section~\ref{subsec:cross}.

\subsection{Auto-only fits}
\label{subsec:auto}

\begin{table}
    \centering
    \begin{tabular}{lccc}
        \hline
        \hline
        Bounds &LRG          &ELG\\
        \hline
        $\log_{10} M_\mathrm{cut}$   & $[12,13.8]$  & $[11.6, 12.6]$\\
        $\log_{10} M_1$              & $[12.5,15.5]$& $[12.5, 18.0]$\\
        $\sigma$                & $[0.0,3.0]$  & $[0.0, 3.0]$\\
        $\alpha$                & $[0.0,1.5]$  & $[0.0, 1.2]$\\
        $\kappa$                & $[0.0,1.0]$ & $[0.0, 10.0]$\\
        $p_\mathrm{max}$        & -           & $[0.05, 1.0]$\\
        $\gamma$                & -           & $[1.0, 15.0]$\\
        \hline
        $\log_{10}M_\mathrm{1,EE}$ & -        & $[12.5, 15.5]$ \\
        $\log_{10}M_\mathrm{1,EL}$ & -        & $[12.5, 35.0]$ \\
        \hline
        $B_\mathrm{cent}$       & $[-1.0,1.0]$ & $[-1.0,1.0]$ \\
        $B_\mathrm{sat}$       & $[-1.0,1.0]$ &  $[-1.0,1.0]$ \\
        \hline        
    \end{tabular}
    \caption{Prior bounds used for LRG and ELG extended HOD model. We adopt the prior on ELG satellite $\alpha$ from Equation~\ref{equ:alpha_prior}, Otherwise, the bounds are chosen to be broad and nonrestrictive. The first 7 rows refer to the vanilla model parameters of the two tracers. The 8-9th rows refer to 1-halo conformity parameters, whereas the last two rows refer to the galaxy assembly bias parameters. Units of mass are in $h^{-1}M_\odot$.}
    \label{tab:priors}
\end{table}

\begin{figure}
    \hspace{-0.0cm}
    \includegraphics[width=0.45\textwidth]{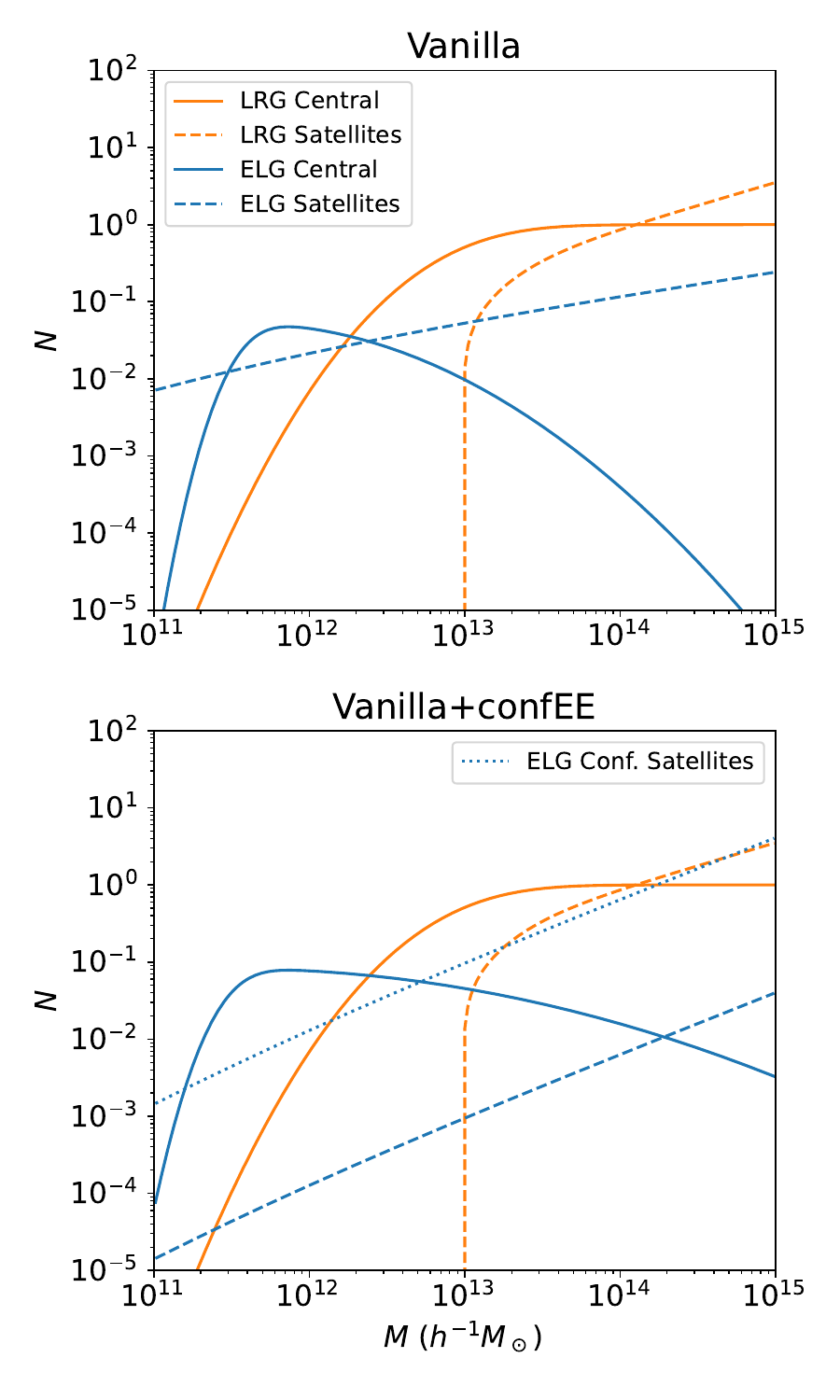}
    \vspace{-0.2cm}
    \caption{The best-fit HOD of the auto-only fits. The top panel shows the best-fit values of the vanilla model, whereas the bottom panel shows the best-fit values of the vanilla+$M_\mathrm{1, EE}$ model. The orange curves denote the HOD of the LRGs while the blue curves denote the HOD of the ELGs. The solid and dashed lines correspond to centrals and satellites, respectively. In the bottom panel, we also show the HOD of the conformal satellites in orange dotted line. The orange dashed line showing the non-conformal satellites goes below the plotted range. }
    \label{fig:hod_auto}
\end{figure}

We first compare several different HOD models by fitting the LRG$\times$LRG and ELG$\times$ELG auto-correlation functions. We summarize the fiducial priors in Table~\ref{tab:priors}, and present the goodness-of-fits and model scores in Table~\ref{tab:aic}. The best-fit parameters of the select models are tabulated in Table~\ref{tab:bestfits}. The best-fit HODs are shown in Figure~\ref{fig:hod_auto}, where the ELG HOD is shown in blue and LRG is shown in orange. The bottom panel shows the best-fit HOD when including conformity, with the dotted line showing the number of satellites in halos with an ELG central and the dashed line showing the number of satellites in halos without an ELG central. We see a very strong conformity effect; the probability of a halo hosting an ELG satellite at fixed halo mass increases by two orders of magnitude when there is a central ELG.

\begin{figure*}
    \hspace{-0.7cm}
    \includegraphics[width=1\textwidth]{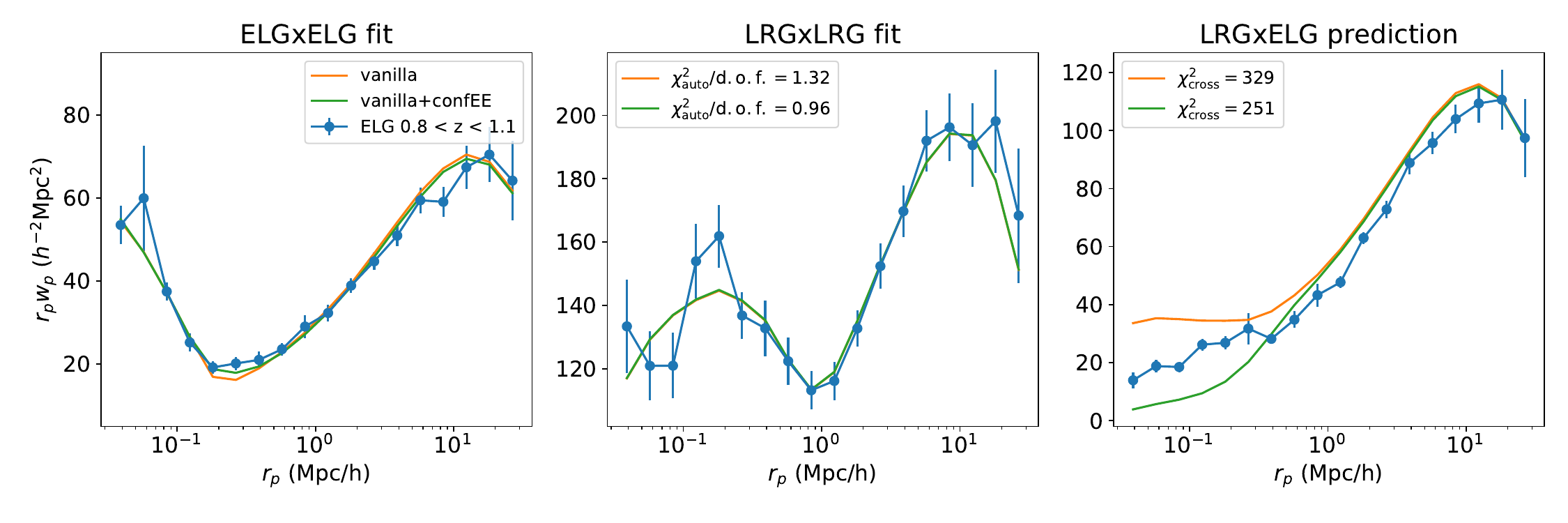}
    \vspace{-0.2cm}
    \caption{Best-fits on the DESI One-Percent auto-correlation functions using the vanilla HOD model and the vanilla+$M_\mathrm{1, EE}$ conformity model. For this comparison, we are only fitting the ELG$\times$ELG and LRG$\times$LRG auto-correlation functions. The LRG$\times$ELG cross-correlations shown for the two models are blind predictions. Also note that we are fitting LRG HOD on the LRG auto-correlation first and then fixing the LRG HOD parameters when fitting ELG HOD parameters. Thus, the orange and green curves in the second panel fully overlaps. The vanilla model achieves a best-fit $\chi^2$/d.o.f. = 1.32 on the auto-correlation functions, whereas the vanilla+$M_\mathrm{1, EE}$ model achieves $\chi^2$/d.o.f. = 0.96. On the cross-correlation function, the vanilla model yields a $\chi^2 = 329$, whereas the vanilla+$M_\mathrm{1, EE}$ model yields $\chi^2 = 251$. 
    }
    \label{fig:wps_nocrossfit}
\end{figure*}

Figure~\ref{fig:wps_nocrossfit} visualizes the best-fit clustering predictions of the vanilla HOD model in orange and the conformity model in green. The legends in the second panel shows the goodness-of-fits of the two models on the auto-correlation functions, whereas the legends in the third panel shows the significance of the difference between the predicted cross-correlation functions and the data. The error bars are computed from 60 jackknife regions.


\begin{table}
    \centering
    \begin{tabular}{l|cc|cc}
        \hline
        \hline
        \multirow{2}{*}{Model} &\multicolumn{2}{c}{Auto-only} &\multicolumn{2}{c}{Auto+Cross}\\
        \cline{2-5}
          & $\chi^2$/d.o.f. & AIC   & $\chi^2$/d.o.f. & AIC \\
        \hline
        Vanilla                                             & 1.32  & 56 & 2.58 & 132\\
        Vanilla + $M_\mathrm{1, EE}$                        & 0.96  & 48 & 2.16 & 114\\
        Vanilla + $M_\mathrm{1, EE}$ + $M_\mathrm{1, EL}$   & 0.95  & 49 & 2.12 & 113\\
        Vanilla + $M_\mathrm{1, EE}$ + $B^\mathrm{LRG}$                          & - & - & 2.15 & 114 \\
        Vanilla + $M_\mathrm{1, EE}$ + $B^\mathrm{ELG}$                          & - & - & 1.86 & 102 \\
        \hline        
    \end{tabular}
    \caption{Goodness-of-fit and model AIC scores for our HOD models. The vanilla HOD model refers to the model described from Equation~\ref{equ:zheng_hod_cent}-Equation~\ref{equ:elg_sat}. $M_\mathrm{1, EE}$ and $M_\mathrm{1, EL}$ refer to 1-halo conformity between ELG satellites and ELG centrals and LRG centrals, respectively. See Equation~\ref{equ:elgconfhod} for descriptions. The $B^\mathrm{LRG}$ and $B^\mathrm{ELG}$ denote the addition of LRG and ELG assembly bias, respectively. See Equation~\ref{equ:AB} for definitions. Note that The LRG assembly bias is parameterized in terms of local over-density, whereas the ELG assembly bias is parameterized in terms of local shear.}
    \label{tab:aic}
\end{table}

We start with vanilla HOD models (``Vanilla'' in Table~\ref{tab:aic}, Equation~\ref{equ:zheng_hod_cent}-Equation~\ref{equ:elg_sat}), with no assembly bias or conformity. 
The vanilla model achieves a reasonably good fit on the auto-correlation functions ($\chi^2$/d.o.f.=1.32) and generates a reasonably consistent prediction of the LRG$\times$ELG cross-correlation function. This is surprising at first glance because from our \tng\ analysis, we expected that a vanilla model without conformity would fail to produce the strong small-scale amplitude of ELG auto-correlation function and the low amplitude of the LRG$\times$ELG cross-correlation function. However, the vanilla model can still produce some of these features via a lowest ELG $\alpha$ value allowed, which assigns a large number of ELG satellite galaxies to low mass halos (see top panel of Figure~\ref{fig:hod_auto}), thus generating excess numbers of ELG pairs with small separation.
This is qualitatively consistent with the findings of \cite{2023Rocher}, where the authors also found a preference for very small and even negative $\alpha$ in an ELG auto-correlation analysis with vanilla HODs. Note that our best-fit $\alpha$ is 0.3 whereas \cite{2023Rocher} found a negative $\alpha$, likely due to the fact that we assign satellites to particles while \cite{2023Rocher} assumed an NFW profile in their vanilla model. 



Next, we conduct the equivalent analyses of the auto-correlation functions with our conformity model (``Vanilla+$M_\mathrm{1, EE}$'', Equation~\ref{equ:elgconfhod}). 
The green curve in Figure~\ref{fig:wps_nocrossfit} shows the best-fit with a vanilla model plus the ELG central-satellite conformity parameter $M_\mathrm{1, EE}$ as defined in Equation~\ref{equ:elgconfhod}. The fit on the two auto-correlation functions achieves an excellent $\chi^2$/d.o.f. of 0.96, a significant improvement over the vanilla model. The improvement comes largely from the 1-halo to 2-halo transition regimes at $r_p\sim 0.3h^{-1}$Mpc, where the conformity model better produces the turnover. 

The right panel of Figure~\ref{fig:wps_nocrossfit} shows the cross-correlation predictions from the two models. Neither predictions agree particularly well with the data, though the conformity model reports a smaller $\chi^2$. On the very small scales, the data sits between the vanilla prediction and the conformity model prediction, but on larger scales both models make consistent predictions that are systematically larger than the data. This suggests that the linear bias of the two tracers are well constrained by the auto-correlations, regardless of whether conformity is included. The systematic difference between the two predictions and the data potentially points to effects such as galaxy assembly bias. As we discussed in section~\ref{subsec:tng_ab}, hydrodynamical models of DESI samples do support the inclusion of galaxy assembly biases in both the LRG and ELG samples. We will further discuss the inclusion of galaxy assembly bias in the following section on cross-correlation fits.

We also test the addition of ELG$\times$LRG cross conformity parameter $M_\mathrm{1, EL}$ and report the goodness-of-fitness and AIC scores in Table~\ref{tab:aic} (``Vanilla+$M_\mathrm{1, EE}$+$M_\mathrm{1, EL}$''). We find that the addition of $M_\mathrm{1, EL}$ does not significantly improve the fit and the model is disfavored by the data compared to the model with just $M_\mathrm{1, EE}$. We find the same conclusion in an Auto+Cross fit so we omit $M_\mathrm{1, EL}$ in the rest of this analysis.



\begin{table*}
    \centering
    {\renewcommand{\arraystretch}{1.5}
    \begin{tabular}{l|l|c|c|c|c|c}
        \hline
        \hline
        \multirow{2}{*}{Tracer} & \multirow{2}{*}{Parameters} & Vanilla & Vanilla + $M_\mathrm{1, EE}$& Vanilla & Vanilla + $M_\mathrm{1, EE}$ & Vanilla + $M_\mathrm{1, EE}$ + $B^\mathrm{ELG}$\\[-2pt]
        & & (Auto) & (Auto) & (Auto+Cross) & (Auto+Cross) & (Auto+Cross)  \\[2pt]
        \hline

        \multirow{9}{*}{LRG} & $\log M_\mathrm{cut}$& 12.98&12.98&12.98&12.98&12.98\\
        
        & $\log M_1$&14.06&14.06&14.06&14.06&14.06\\

        & $\sigma$&0.40&0.40&0.40&0.40&0.40\\
        
        & $\alpha$&0.59&0.59&0.59&0.59&0.59\\
        
        & $\kappa$&1.00&1.00&1.00&1.00&1.00\\[2pt]

        \cline{2-7}
        
        & $f_\mathrm{sat}$&0.097&0.097&0.097&0.097&0.097\\
        
        & $\log_{10} \overline{M}$&13.22&13.22&13.22&13.22&13.22\\[2pt]

        \cline{2-7}       

        \hline

        \multirow{12}{*}{ELG} & $\log M_\mathrm{cut}$ & 11.58 & 11.50 & 11.50 & 11.50 & 11.64\\
        
                              & $\log M_1$            & 16.98 & 16.75 & 17.97 & 16.12 & 15.17\\

                              & $\sigma$              & 0.77  & 1.37  & 0.75  & 0.42  & 1.04\\
        
                              & $\alpha$              & 0.31  & 0.79  & 0.30  & 0.48  & 0.70\\
        
                              & $\kappa$              & 0.34  & 0.33  & 0.11  & 1.30  & 0.18\\
        
                              & $p_\mathrm{max}$      & 0.06  & 0.15  & 0.07  & 0.06  & 0.33\\
                              & $\gamma$              & 5.17  & 8.62  & 5.37  & 4.78  & 5.60\\[2pt]

        \cline{2-7}
                              & $\log_{10}M_\mathrm{1, EE} $& - & 14.21 & -   & 14.46 & 14.44\\
        
        & $B_\mathrm{cent}$&-&-&-&-&0.14\\
        
        & $B_\mathrm{sat}$&-&-&-&-&-0.42\\[2pt]
        
        \cline{2-7}       

        & $f_\mathrm{sat}$&0.525&0.023&0.314&0.107&0.102\\
        
        & $\log_{10} \overline{M}$&11.98&12.07&11.91&11.98&12.23\\[2pt]
        \hline
    \end{tabular}%
    }
    \caption{LRG and ELG best-fit parameters. We also quote the marginalized satellite fraction $f_\mathrm{sat}$ and the sample completeness $f_\mathrm{ic}$. Units of mass are given in $h^{-1}M_\odot$.
    }
    \label{tab:bestfits}
\end{table*}
  
\subsection{Auto+cross fits}
\label{subsec:cross}

\begin{figure}
    \hspace{0cm}
    \includegraphics[width=0.45\textwidth]{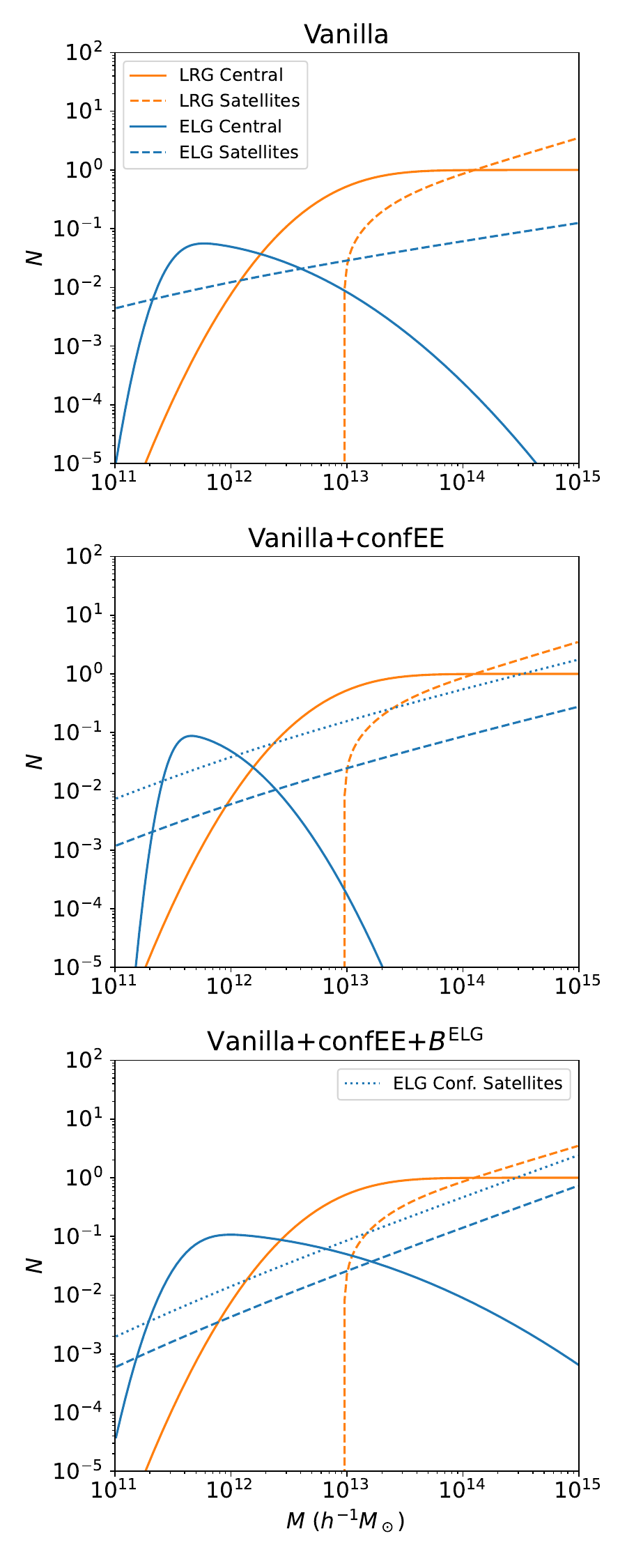}
    \vspace{-0.2cm}
    \caption{The best-fit HOD of the auto+cross fits. The top to bottom panels show the best-fit values of the vanilla model, the vanilla+$M_\mathrm{1, EE}$ model, and the vanilla+$M_\mathrm{1, EE}$+$B^\mathrm{ELG}$ models, respectively. The orange curves denote the HOD of the LRGs while the blue curves denote the HOD of the ELGs. The solid and dashed lines correspond to centrals and satellites, respectively. The dotted blue lines show the HOD of the conformal satellites. }
    \label{fig:hod_cross}
\end{figure}

\begin{figure*}
    \hspace{-0.7cm}
    \includegraphics[width=1\textwidth]{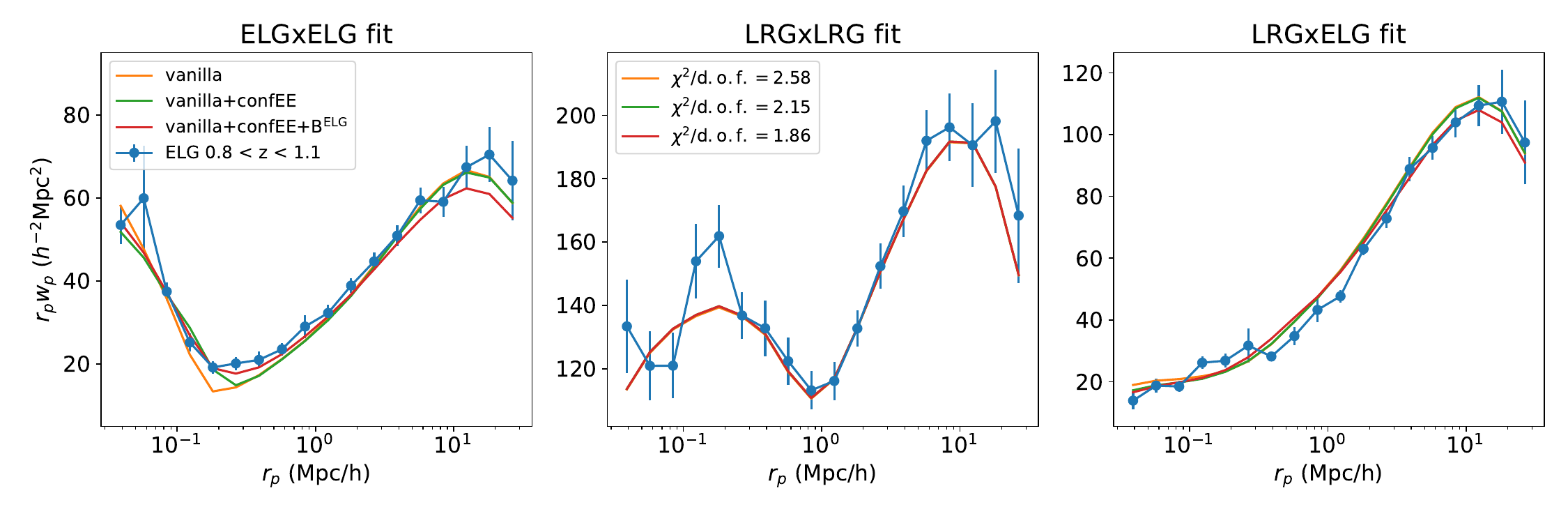}
    \vspace{-0.2cm}
    \caption{Best-fits on the DESI One-Percent auto and cross-correlation functions using the vanilla HOD model, the vanilla+$M_\mathrm{1, EE}$ model, and the vanilla+$M_\mathrm{1, EE}$+$B^\mathrm{ELG}$ models. For this comparison, we are fitting the all three auto and cross-correlation functions. 
    }
    \label{fig:wps_crossfit}
\end{figure*}

Similar to the auto-correlation analysis in the previous sub-section, we apply our different HOD models to the auto+cross correlation fits. The goodness-of-fits and model evidence are presented in the last columns of Table~\ref{tab:aic}. The best-fit HODs are shown in Figure~\ref{fig:hod_cross}, and the best-fit parameters are summarized in Table~\ref{tab:bestfits}. The best-fit predicted clustering are shown in Figure~\ref{fig:wps_crossfit}. 

Immediately, the auto+cross fits confirm the key findings of the auto-only analysis in terms of 1-halo conformity. We find that the inclusion of the conformity parameter $M_{1,\mathrm{EE}}$ continues to be favored by data with a significant improvement in $\chi^2$/d.o.f., whereas the addition of $M_{1,\mathrm{EL}}$ parameter is not favored. Visually, both the ``vanilla+confEE'' model and the vanilla model continue to produce reasonably good fits on the data. Again the vanilla model can roughly produce the small-scale features with a low $\alpha$, but it struggles to simultaneously produce the shape of the ELG auto-correlation upturn and the ELG$\times$LRG correlation downturn. The 1-halo to 2-halo transition regime in the ELG auto-correlation function continues to be the most visually discernible difference in the model predictions. 

We also test the inclusion of galaxy assembly bias features in the model. Following preceding studies with hydrodynamical simulations and semi-analytic models, we parameterize LRG assembly bias in terms of local scalar environment, whereas we parameterize ELG assembly bias in terms local shear (see section~\ref{subsec:tng_ab} for details). As shown in Table~\ref{tab:aic} under model names ``Vanilla+$M_\mathrm{1, EE}$+$B^\mathrm{LRG}$'' and ``Vanilla+$M_\mathrm{1, EE}$+$B^\mathrm{ELG}$'', the addition of LRG assembly bias does not significantly improve the fit on clustering, but the addition of ELG assembly bias does. This result is significant as the first indirect evidence for ELG assembly bias from data. Interpreting the best-fit values in Table~\ref{tab:bestfits}, the ELG centrals slightly prefer lower shear environments, whereas the satellites prefer higher shear environments. 

The combination of a strong central-satellite conformity signal and a preference for high-shear environments in the satellites is again consistent with the picture that the ELGs preferentially select tidally disturbed galaxies undergoing mergers and displaying triggered star formation at the time of observation. We do not yet know the significance of our assembly bias constraints without proper posterior sampling. We reserve a more comprehensive exploration of ELG assembly bias for a future analysis. 

An interesting discrepancy in these fits is in the ELG satellite fraction. While the vanilla model predicts a rather large ELG satellite fraction of 30--$50\%$, the conformity model produces a low satellite fraction at 2--10$\%$. In comparison, for eBOSS ELGs, \cite{2019Guo} found a satellite fraction of 13--17$\%$, whereas \citet{2016Favole} found a satellite fraction of $22.5\pm2.5\%$. These studies used vanilla models without conformity as eBOSS clustering measurements were not sufficiently precise on the very small scales to detect potential signatures of conformity. In more recent DESI analyses, \cite{2023Rocher} found a similar satellite fraction of $2.3\%$ when including maximal conformity but assuming a slightly different vanilla model. \cite{2023Yu} analyzed the same data with a sub-halo abundance matching model and also found a low satellite fraction for ELGs, at 3-5$\%$ depending on the specifics of the model. However, \cite{2023Gao} also used a SHAM model to predict the auto- and cross-correlation functions and found a satellite fraction of $\sim 15\%$, higher than the other DESI analyses. These comparisons suggest that a combination of features specific to the DESI sample and the inclusion of conformity in the modeling potentially contributed to a low satellite fraction, but there are still some disagreements between the different methods.

It makes sense that conformity reduces the predicted satellite fraction as it accounts for the excess small-separation pairs through tight central-satellite pairs, whereas the vanilla model produces the small-scale signature by producing large numbers of satellites in small halos by preferring a low $\alpha$. One caveat is that a low $\alpha$ produces satellites in small halos without centrals, which cannot be effectively distinguished from centrals with clustering alone. In other words, we can simply re-label satellites without centrals as centrals and it would     result in the same clustering prediction, or that the satellite fraction is somewhat degenerate with the definition of centrals and satellites in these small halos. \cite{2020Avila} found a similar conclusion with eBOSS data that the inferred ELG satellite fraction varies with the assumed HOD model, anywhere between 20-50$\%$ for their models.

\bigskip

Overall, we find some statistical evidence that the inclusion of conformity and galaxy assembly bias in the ELG HOD are favored by by the observed auto and cross-correlation functions. This is consistent with the physical picture of triggered starburst we proposed in section~\ref{sec:cosmos} and the conformity signal we find in \tng\ in section~\ref{sec:tng}. However, we continue to find that the vanilla model can produce a reasonably good fit to the clustering by generating a large number of ELG satellites in small halos with a low $\alpha$, and that the improvement when including conformity and galaxy assembly bias is not particularly visually obvious in the clustering comparisons. Thus, a more rigorous statistical comparison of the models are needed to quantify the model evidences. Our results are meaningful as part of a broader push on ELG modeling, and we reserve a more complete HOD analysis for the future. 

Our model also ultimately did not produce $\chi^2$/d.o.f.~$\approx 1$ on the auto+cross fit. This could be due to the fact that we do not consider the full covariance matrix for this analysis due to limited number of jackknife regions. We speculate that positive off-diagonal terms would reduce the significance of model discrepancies and reduce the overall $\chi^2$. On the modeling side, we have ignored redshift evolution, which can prove important as we push towards $\chi^2$/d.o.f.~$= 1$. 

There is also the possibility of additional motivated extensions to the HOD model. \cite{2023Rocher} proposed the inclusion of an extended halo profile. Specifically, their default model assigned ELG satellite galaxies according to an NFW profile, but they found that a modified NFW profile allowing satellite galaxies to be assigned further away from the halo center is significantly reduced the $\chi^2$. The extended profile is motivated by the notion that star-forming galaxies are quickly quenched when they fall deep into the halo potential well. There are several lines of observational evidence showing the quenched fraction of galaxies is indeed radially dependent within a halo in SDSS \citep{2012Wetzel, 2007Blanton}, but the connection between those observations and high redshift ELGs is unclear and we did not find evidence of deviations from NFW profile in ELGs in hydrodynamical simulations \citep{2022mHadzhiyska, 2022Yuan}. We suggest a detailed study of ELG radial distributions in the future. 

\section{Constraints from cluster cross-correlations}
\label{sec:redmapper}

Given the degeneracy between conformity and low satellite occupation power-law index $\alpha$ seen in this analysis and in \cite{2023Rocher}, observables that can direct constrain $\alpha$ become particularly interesting. In this section, we present one such observable, where we tabulate DESI ELG satellite counts around known galaxy clusters. Combined with the independently measured cluster richness-mass relationship, this gives us a direct way to constrain the ELG satellite $\alpha$ in massive halos and potentially break this degeneracy. 

We use the Legacy Survey redMaPPer cluster catalog \citep{2020Chitham}. redMaPPer is a red-sequence photometric cluster finding algorithm that identifies galaxy clusters as over-densities of red galaxies, and estimates the probability that each red galaxy is a cluster member following a matched filter approach that models the galaxy distribution as the sum of a cluster and background component. A richness parameter $\lambda$ is computed for each cluster and is defined the sum total of the membership probabilities of all the galaxies. It has been found that richness $\lambda$ tightly correlates with the total mass of the cluster \citep[recent examples include][]{2019McClintock, 2017Simet, 2016Baxter}, thus giving us a way to directly relate ELG occupation to halo mass. 

The Legacy Survey redMaPPer cluster catalog is obtained by running redMaPPer on DR9 photometry from the DESI Legacy Imaging Surveys \citep{2019Dey}, covering approximately 14,300 deg$^2$. For our analysis, we select 2490 clusters that are fully within the DESI SV3 footprint and have spectroscopically confirmed Bright Central Galaxies (BCGs) in DESI. Specifically, we only consider clusters whose centers are within the SV3 footprint and their distances to the edge of footprint is at least 0.22 degrees, which translates to a comoving distance of 9.4 $h^{-1}$Mpc at $z = 1$. This procedure removes the need to consider boundary effects. We use DESI spectroscopic redshift measurements (spec-$z$) for the BCGs as the cluster redshifts to partially remove the need to consider cluster redshift uncertainty and associated projection effects. We further remove clusters with richness $\lambda < 5$ to reduce our exposure to systematics due to richness bias, resulting in 1447 clusters for our final sample. 

To compute the ELG satellite count as a function of cluster mass, we first search for ELG neighbors in 2D around each cluster in a radius that is 3 times the default richness-dependent redMaPPer radius. The number 3 is chosen to roughly reproduce the expected $R_{200\mathrm{m}}$ around a $10^{14}\ h^{-1}M_\odot$ halo with richness $\sim 10$. We only select ELG neighbors within a finite cylinder along the line-of-sight, centered on the BCG spec-$z$. Specifically, for each cluster with spectroscopic redshift $z_\mathrm{BCG}$, we only include ELGs within redshift range $z_\mathrm{BCG}\pm 3(\sigma_\mathrm{v}/c)$, where $\sigma_\mathrm{v}$ is the cluster velocity dispersion. The velocity dispersion measurement is fairly uncertain, so we set the velocity dispersion measurements to both 500km/s and 1000km/s, and we find that the two different values result in a $\sim 1\%$ change in the ELG satellite counts. Considering that we are using spec-$z$ for clusters and the robustness of satellite counts to the choice of velocity dispersion, our results should be robust to projection effects in the cluster--ELG pair counts. 

\begin{figure}
    \hspace{-0.2cm}
    \includegraphics[width=0.48\textwidth]{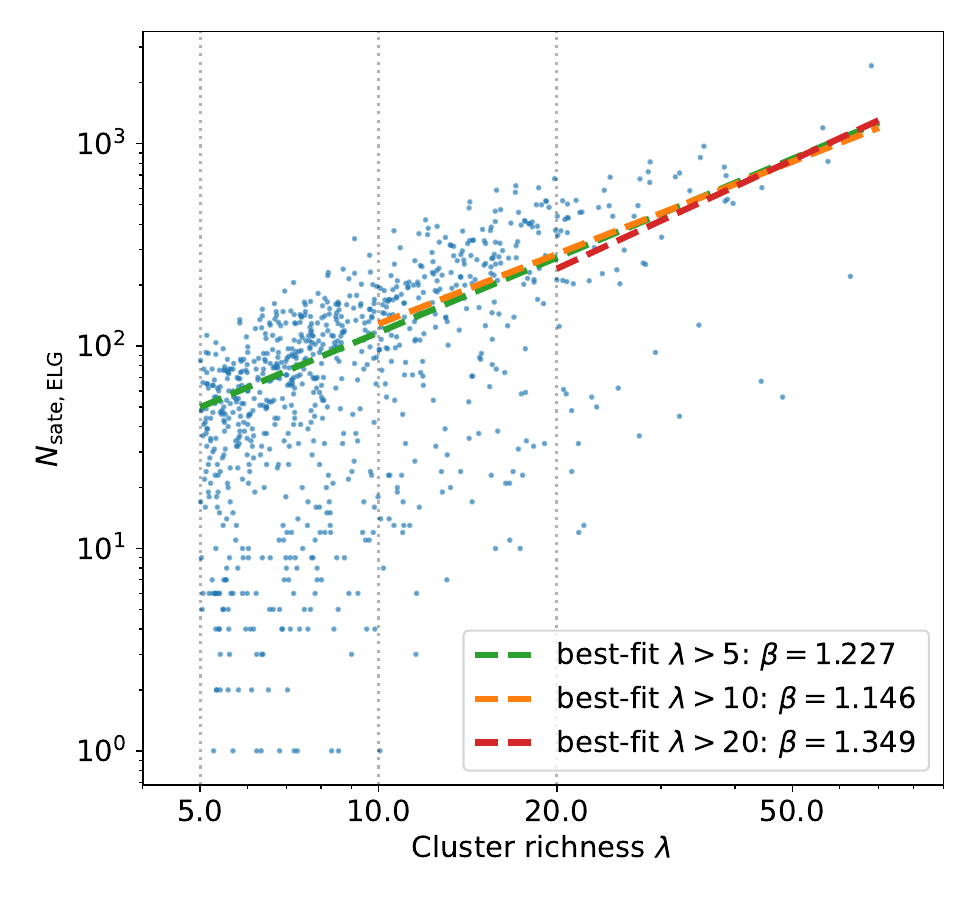}
    \vspace{-0.2cm}
    \caption{The number of DESI ELG satellites per redMaPPer cluster as a function of cluster richness. The blue points show the individual clusters. The three dashed lines show the power law fits through clusters with different minimum richness cuts. The power law slope measurement varies by $\sim 20\%$. The statistical uncertainty on the best-fit $\beta$ is below 0.001 in all three cases. }
    \label{fig:cluster_alpha}
\end{figure}
Figure~\ref{fig:cluster_alpha} shows the number of ELG satellites as a function of cluster richness. The blue points showcase the sample distribution. The dashed lines show the best-fit power law with increasing minimum $\lambda$. Clearly, we see an monotonically increasing trend, suggesting higher richness clusters (more massive halos) host more ELG satellites. This strongly rejects a negative ELG $\alpha$, in which case less massive halos host more ELG satellites. 
To further quantify this measurement, we define the satellite-occupation vs. richness correlation parameter $\beta$, as in
\begin{equation}
    N_\mathrm{sate,ELG}\sim \lambda^\beta.
\end{equation}
The dashed green line shows the fit to the full cluster sample ($\lambda > 5$), resulting in best-fit $\beta = 1.23$.

However, the optical richness of redMaPPer clusters can be subject to large biases due to projection effects in cluster member assignments \citep{2015Rozo, 2019aCostanzi, 2018Sohn, 2020Sunayama}. \cite{2021Myles} showed that this bias can be as large as $20\%$ and increases with lower richness and higher redshift. Unfortunately, we do not currently have measurements of this bias for clusters at $z\sim 1$. For the purposes of this fairly qualitative analysis, we conduct an internal consistency check by measuring $\beta$ for progressively larger minimum richness. These measurements are shown with dashed orange and red lines in Figure~\ref{fig:cluster_alpha}. The best-fit $\beta$ is self-consistent at $\pm 10\%$, varying between 1.15 and 1.35. This potentially demonstrates that our $\beta$ measurements are not strongly affected by richness bias, especially considering $\lambda > 20$ was employed for the fiducial cosmology analysis in \cite{2020Abbott} and \cite{2021Costanzi}. Nevertheless, we acknowledge that with current data it is not currently feasible to explicitly calculate richness bias for our sample and that the richness values used in our analysis can still be significantly affected by projection effects. 

To translate constraints on $\beta$ to $\alpha$ in the HOD, we need to know the power law index in the cluster mass-richness relationship. For redMaPPer clusters, recent studies have used clustering and weak lensing to directly constrain cluster masses and derive mass-richness relationships. At lower redshift $z < 0.33$, \cite{2017Simet} derived $M\sim \lambda^{1.3\pm 0.1}$ from weak lensing measurements, whereas \cite{2016Baxter} derived $M\sim \lambda^{1.2\pm 0.2}$ from clustering measurements. \cite{2019McClintock} uses DES Y1 imaging to calibrate redMaPPer mass-richness relationship up to $z = 0.65$, deriving $M\sim \lambda^{1.35\pm 0.13}$. \cite{2019Murata} used HSC imaging to study the redshift evolution of the mass-richness relationship of the CAMIRA clusters and found the power law index to be insensitive to redshift at $z < 1$. Combining these results, we suggest a conservative mass-richness relationship of 
\begin{equation}
    \lambda \sim M^{0.7-0.9}.
    \label{equ:mass-richness}
\end{equation}
Combining Equation~\ref{equ:mass-richness} and our range for $\beta $, we arrive a range for the satellite power index $\alpha$ that is 
\begin{equation}
    0.8<\alpha < 1.2.
    \label{equ:alpha_prior}
\end{equation}

This range is broadly consistent with recent MilleniumTNG studies and semi-analytic models \citep{2022mHadzhiyska, 2022Yuan, 2020Avila}. 
This range favors models with higher inferred $\alpha$, such as the Vanilla+$M_\mathrm{1, EE}$ (auto) and Vanilla+$M_\mathrm{1, EE}$+$B^\mathrm{ELG}$ (cross) models in Table~\ref{tab:bestfits}. It disfavors vanilla models that require a much lower $\alpha = 0.3$ to describe the clustering. Moreover, this measurement strongly disfavors the vanilla models in \cite{2023Rocher} that require a negative $\alpha$. 

Finally, we re-iterate that while we have attempted to minimize systematic biases in this exercise by conducting self-consistency tests and employing spectroscopic redshifts whenever possible, our results can still be affected by a range of systematics related to optical richness including projection effects and mis-centering. Nevertheless, this result is important as a first step in breaking the degeneracy between low $\alpha$ and conformity models. 

\section{Producing Conformity with UniverseMachine}
\label{sec:um}
So far, we have motivated the inclusion of conformity through both hydrodynamical models and observations. We have also used COSMOS data to build a physical picture for the ELGs that should naturally generate conformity. In this section, we explore the origin of ELG conformity through a set of tests with UniverseMachine.

Specifically, we stress test the UniverseMachine model by varying its parameters to see if it can consistently predict the DESI ELG stellar mass function, SFR distribution, and clustering. Our tests challenge a fundamental assumption in UniverseMachine that ties galaxy SFR exclusively to halo mass and halo mass assembly history. We argue that the type of conformity signal we see requires a more flexible star formation model that includes additional merger-triggered starburst that is independent of mass assembly. These tests also further distinguish merger-triggered conformity from galaxy assembly bias. 

\subsection{UniverseMachine model}
\textsc{UniverseMachine}~\citep[UM;][]{2019MNRAS.488.3143B} is an empirical galaxy--halo connection model that predicts galaxy star formation rates from halo mass and halo assembly histories. It is a flexible framework that models the full evolution histories of galaxies anchored on dark matter halo merger trees from cosmological simulations, and it is simultaneously constrained by observed galaxy stellar mass functions, UV luminosity functions, quenched fractions, cosmic star formation history, and galaxy clustering over a wide range of galaxy mass and redshifts (up to $z\sim 8$). The galaxy clustering constraints included in \citet{2019MNRAS.488.3143B} were from SDSS DR7~\citep{2003MNRAS.341...33K,2004MNRAS.351.1151B,2009ApJS..182..543A} that cover auto and cross correlations of all, quenched, and star-forming galaxies with $M_{\ast} > 10^{10.3}\mathrm{M_{\odot}}$ at $z<0.7$.

We briefly summarize the UM model framework. The model uses peak halo maximum circular velocity, $v_{\rm Mpeak}$, as a halo mass proxy that is both resistant to pseudo evolution and consistent for either satellites and centrals. The model parameterizes the redshift-dependent mean star formation rate and galaxy quenched fractions ($f_Q$) as a function of $v_{\rm Mpeak}$ (with some parameterized intrinsic scatter). These parameterizations determine the distribution of galaxy SFRs in each redshift and halo mass bin (double Gaussian normalization ratio given by $f_Q$). In each halo mass ($v_{\rm Mpeak}$) bin, the model tends to map higher SFR onto earlier-forming halos with lower recent accretion rates, capturing the effect of halo assembly bias. The halo recent accretion rate at each redshift ($z$) is defined as~\citep{2019MNRAS.488.3143B}:
\begin{equation}
\label{eq:dvm}
    \Delta v_{\rm max}(z) = \frac{v_{\rm max} (z)}{v_{\rm max}\left(\max (z_{\rm dyn}, z_{\rm Mpeak})\right)} ,
\end{equation}
with $z_{\rm dyn}$ being the redshift one dynamical timescale ago from $z$ and $z_{\rm Mpeak}$ being the redshift of peak halo mass. For each halo, its SFR is determined by a percentile rank function:
\begin{equation}
    C({\rm SFR}) = C\left(r_c C^{-1}(C(\Delta v_{\rm max})) + \sqrt{1-r_c^2} R(t) \right),
    \label{equ:C_sfr}
\end{equation}
where $C(x) = 0.5 + 0.5 {\rm erf} (x/\sqrt{2})$ is the cumulative percentile rank function. $r_c$ is the rank correlation coefficient between the percentile rank in SFR and percentile rank in $\Delta v_{\rm max}$ in each halo $v_{\rm Mpeak}$ bin (dependent on $v_{\rm Mpeak}$ and $z$).
$r_c$ controls the strength of the correlation between SFR and halo accretion, while the random series $R(t)$ contributes a time-correlated piece. The random time series is parameterized as:
\begin{equation}
    R(t) = f_{\rm short} S_{\rm short} (t) + \sqrt{1-f_{\rm short}^2} S_{\rm long}(t),
\end{equation}
where $S_{\rm short}(t)$ is an uncorrelated unit random variable that models the short-timescale stochasticity of the SFR. $S_{\rm long}(t)$ is a correlated random Gaussian series with correlation timescale set to the dynamical timescale (correlation coefficient for two time steps separated by $\delta t$ is $\exp (-\delta t/\tau_{\rm dyn})$). Finally, $f_{\rm short}$ is the parameter that models the relative proportion of short versus long timescale variations of SFR. 

\subsection{Mimicking ELG conformity}

Within the existing UM model, $r_c$ and $f_{\rm short}$ fully control the assignment of SFR at given halo mass and given  set of halo accretion rates. Hence these two parameters should fully determine any emergent galaxy assembly bias and conformity. For example, larger $r_c$ means stronger correlation between galaxy SFR and halo assembly history at fixed halo mass, resulting in strong galaxy assembly bias by definition. We also expect a large $r_c$ to result in stronger conformity because of the correlated assembly histories of host halo and its subhalos. However, based on the analyses in this paper, we postulate that 1-halo conformity can be potentially attributed to merger-triggered starburst in addition to correlated halo accretion. A SFR model that purely depends on halo mass and recent accretion would not predict the additional star formation triggered by galaxy mergers, which requires an additional dependency on local tidal field or proximity to other galaxies. Thus, by flexing the UM model parameters and testing whether the current UM model space can produce the type of conformity signal we see in DESI, we can determine whether the correlated halo accretion alone can explain the conformity signal we see in DESI and elucidate the distinction between ELG conformity and galaxy assembly bias. 

\begin{figure}
    \hspace{-0.2cm}
    \includegraphics[width=0.5\textwidth]{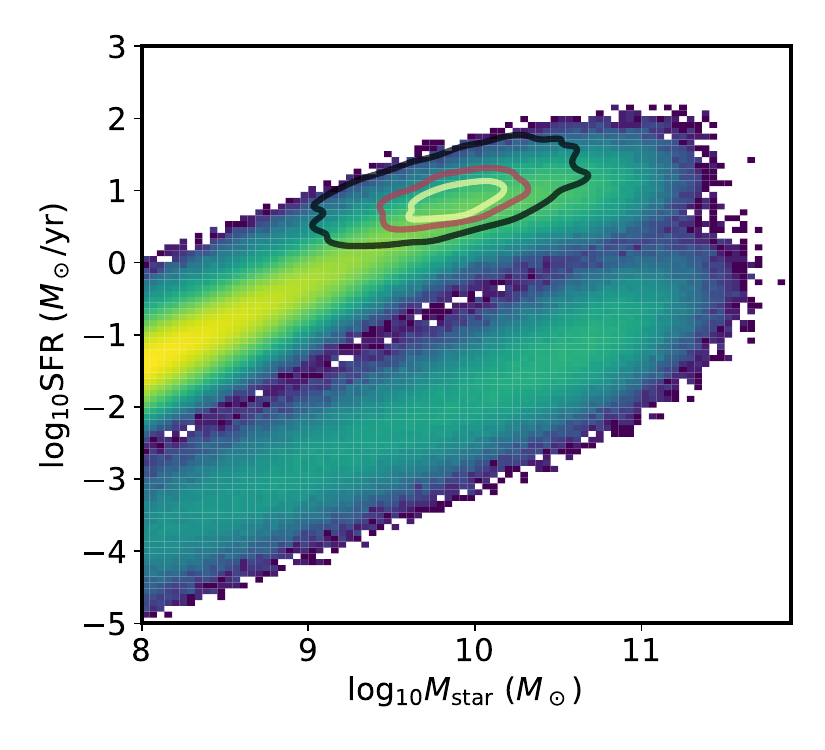}
    \vspace{-0.2cm}
    \caption{The distribution of UniverseMachine galaxies in the stellar mass and star formation rate plane, compared to the mock ELGs shown in contours. We have applied a shift of -0.2 dex to $\log_{10}$SFR and 0.2 dex to $\log_{10}M_\mathrm{star}$ of the COSMOS ELGs to better match the UM stellar mass vs SFR distribution. }
    \label{fig:um_sm_sfr}
\end{figure}

We run three UM models: (1) The fiducial model with default parameters reported in \cite{2019Behroozi} ($r_c = 0.5, f_\mathrm{short} = 1$); (2) A model where we set $f_\mathrm{short} = 0.5$ to reduce the proportion of short timescale variations in SFR, i.e. making the star formation history more correlated on long time scales; (3) A model where we set $r_c = 1$ to obtain maximum correlation between SFR and assembly history. Comparing model (2) and (3) to the fiducial model (1) allows us to test which aspect, if any, of the SFR model can be tweaked to produce the ELG clustering we observe. We run all three models on a 250$h^{-1}$Mpc N-body simulation box with 2560$^3$ particles and particle mass $7.4\times 10^{7}h^{-1}M_\odot$. The simulation is run with \textsc{L-GADGET2} code at a fiducial $\Lambda$CDM cosmology of $\Omega_m = 0.286$, $\Omega_m = 0.047$, $\sigma_8 = 0.82$, $n_s = 0.96$, $h = 0.7$ \citep{2021bWang, 2015Becker, 2015Mao, 2005Springel}. We use the $z = 0.8$ snapshot.

We select ELGs in the UM boxes by directly matching the 2D stellar mass vs. SFR distribution of DESI ELGs in COSMOS (Figure~\ref{fig:completeness2d}). The 2D histograms in Figure~\ref{fig:um_sm_sfr} show the stellar mass and SFR distribution of the full UM catalog at $z = 0.8$. The contours visualize the selected ELGs. We have shifted the COSMOS SFR and stellar mass by -0.2 dex and 0.2 dex, respectively, to better match the values in UM. Such systematic shifts could be necessary due to differences in mass definition or systematics in masses inferred from SED fitting, but the point is that the ELGs are located in qualitatively the same region of parameter space as in Figure~\ref{fig:completeness2d}. We repeat the same selection over the three UM models. Also note that because we are explicitly matching the stellar mass and SFR distribution in the mock ELG selection, the only observable we need to match when we vary the UM parameters is the clustering. 

\begin{figure*}
    \hspace{-0.7cm}
    \includegraphics[width=1\textwidth]{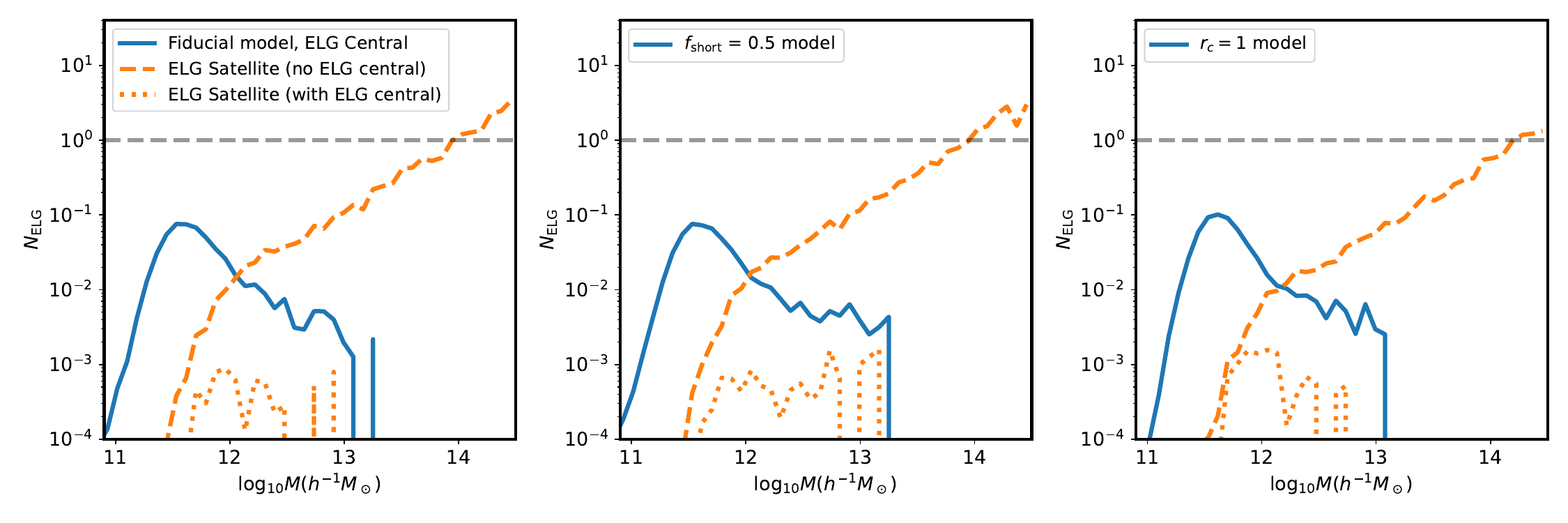}
    \vspace{-0.2cm}
    \caption{HODs of UniverseMachine ELGs selected by sampling the ELG stellar mass function and star formation rate distribution in COSMOS. The three panels correspond to the fiducial model, the model where we set $f_\mathrm{short} = 0.5$, and the model where we set $r_c = 1$. The blue solid lines show the central HOD, whereas the satellites are shown in orange. We separate the satellites into satellites in halos with ELG centrals (in dotted), and satellites in halos without ELG centrals (in dashed). None of the three models generated as much conformity as we saw in \tng\ or in the DESI analysis.  }
    \label{fig:um_hod}
\end{figure*}

Figure~\ref{fig:um_hod} displays the HODs of the selected mock ELGs in the three UM models. The solid blue curves show the HOD of the centrals, whereas the orange curves show the HOD of the ELG satellites. Upon first glance, the three different models produce remarkably consistent ELG HODs that are also qualitatively consistent with our inferred HODs from DESI (Figure~\ref{fig:hod_auto} and Figure~\ref{fig:hod_cross}) and also the \tng\ prediction in Figure~\ref{fig:hod_elg}. The satellite HODs report a power index of $\alpha\sim 0.9$, somewhat larger than the prior range we set from redMaPPer analysis (Equation~\ref{equ:alpha_prior}), but nonetheless disfavors the vanilla HOD models that infer very low $\alpha$ values. The dotted and dashed lines show the HOD of ELG satellites in halos with and without and ELG centrals. Clearly, none of the three models produce a large number of conformity, unlike in \tng\ and we inferred from data. The $r_c = 1$ model produces a larger fraction of satellites with ELG centrals at $7\%$ as opposed to $2\%$ in the fiducial and $f_\mathrm{short} = 0.5$ models. This makes sense as $r_c = 1$ maximally ties SFR to recent accretion history, resulting in some conformity effect. Nevertheless, the weak conformity across the three models suggests that SFR correlated halo accretion as modeled in UM creates a small but insufficient amount of conformity. 

\begin{figure*}
    \hspace{-0.7cm}
    \includegraphics[width=1\textwidth]{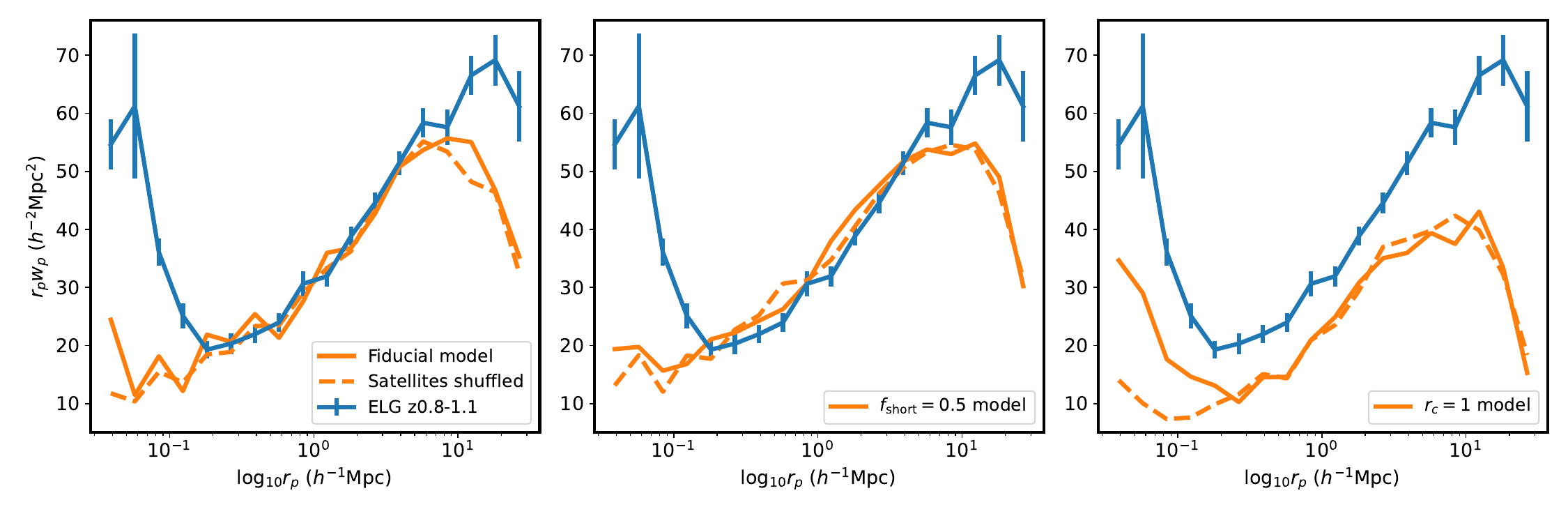}
    \vspace{-0.2cm}
    \caption{The projected 2-point auto-correlation function of UniverseMachine ELGs (in orange) compared to the DESI measurement (in blue). The three panels correspond to different UniverseMachine models. The dashed lines show the clustering when we shuffle satellites amongst halos of the same mass, removing 1-halo conformity. Only the $r_c = 1$ model in the last panel produced a noticeable conformity signature. }
    \label{fig:um_wp}
\end{figure*}

We find the same conclusion when looking at the clustering signatures. Figure~\ref{fig:um_wp} showcases the predicted ELG 2PCF from our UM ELG selection compared to the DESI measurement. The three panels correspond to the three different UM models. We also show in dashed curves the 2PCF prediction when we shuffle satellite galaxies among halos of the same mass, which removes any central-satellite conformity. 

In the left panel showing the fiducial model prediction, we match the observed clustering between 0.2-10$h^{-1}$Mpc. At larger $r_p$, the deviation likely comes from sample variance and missing large scale modes. At smaller $r_p$, the fiducial UM model does not predict the upturn that is indicative of conformity. The middle panel shows that reducing short time scale uncorrelated fluctuation in the star formation history does not have a significant effect on the projected 2PCF. The right panel shows that increasing $r_c$ to 1 does produce some excess clustering on small scales, though not strong as in data. The dashed line shows that the upturn is indeed due to central-satellite conformity. However, setting $r_c = 1$ breaks the consistency in clustering amplitude on large scales. This is not surprising as $r_c = 1$ maximally associates star formation with recent growth, generating a large amount of galaxy assembly bias as result, which modulates the galaxy clustering on 2-halo scales. Because the large scale clustering in UM is well constrained with existing data, a $r_c = 1$ model that significantly changes the large-scale clustering predictions is strongly disfavored.

Before we move on, we point out that it is possible that our ELG selection in UM is not realistic and that an actual SED based selection would report significantly different clustering on 1-halo scales. However, we find that a stellar mass and SFR based selection on \tng\ still results in a strong conformity signal (we omit that discussion for brevity). Similarly, we tested a range of stellar mass and SFR based selection for ELGs in UM, and failed to find a selection that produces significant conformity while also predicting the right number density and stellar mass function. We believe that the lack of conformity in UM cannot be entirely due to selection, but we advocate for a more thorough examination in future work. 

To summarize, while we could generate some conformity signal by maximally correlating SFR with recent accretion history, it does so by breaking consistency in the ELG 2PCF on 2-halo scales. This suggests that the UM model is not currently flexible enough to describe the ELG sample down to 1-halo scales. This makes sense given our physical picture of ELG star formation being at least partially triggered by tidal interactions, whereas UM only correlates star formation with mass assembly, not environmental triggers. Modeling triggered star formation in UM would essentially require modifying Equation~\ref{equ:C_sfr} by adding an association between SFR with local tidal field or the existence and types of neighboring galaxies. We intend to implement such a model extension in the future. 


\section{Discussion}
\label{sec:discuss}

In this paper, we have pushed on our understanding of ELGs and argued for the inclusion of conformity through both observations and physical models.
To complement our results, we survey the literature for other studies of ELG conformity and compare with our analysis. We also examine alternative theories on the origin of conformity.


\subsection{ELG conformity in the literature}
This paper expands significantly on the analysis of \cite{2023Rocher}, by leveraging cross-correlations and external datasets to build a clearer physical picture of the DESI ELG sample. \cite{2023Rocher} did not find a clear preference for conformity in terms of ELG auto-correlation data alone because of a strong degeneracy between conformity and small $\alpha$. Specifically they found that an HOD with conformity results in an $\alpha$ of 0.7-0.9, whereas the model without conformity favors a value of $\alpha = -0.26$. We presented several pieces of evidence that disfavor a negative $\alpha$, including our tests with \tng\ and UM, and the cross-correlation of ELG satellites with redMaPPer clusters. 

\cite{gao_conf} analyzed the DESI LRG$\times$ELG measurements in a subhalo abundance matching (SHAM) framework instead of an HOD. They similarly found that the vanilla SHAM model fails to predict the small-scale clustering of ELGs, and that an extended SHAM model with 1-halo conformity is strongly preferred by the data. This is highly complimentary to our study by showing that the preference for conformity is not simply due to potential limitations of the HOD, but a feature that requires a physical explanation. However, the same paper does not find a need for galaxy assembly bias in order to describe the auto+cross correlation functions. This is likely because SHAM already incorporates at least some galaxy assembly bias, as subhalos readily encode halo assembly information \citep{2016Chaves}. 

There is also evidence for the need of explicit conformity model in an HOD framework before DESI. \cite{2009Ross} analyzed the small-scale clustering of red and blue galaxies in SDSS DR5 and found that a vanilla model that fully mixes the blue and red galaxies does not fit the data, whereas a model that segregates blue and red galaxies into distinct halos (akin to conformity) does. Additional evidence of conformity in red vs blue galaxies was also seen in the Deep Extragalactic Evolution Probe (DEEP2) spectroscopic survey \citep{2013Newman}. Specifically, \cite{2008Coil} showed that the projected clustering of blue galaxies at $0.7 < z < 1.2$ shows a strong upturn on very small scales, which is not produced by a vanilla HOD fit \citep{2010Tinker}. While these studies did not attempt a conformity model, the clustering signatures are fully consistent with what we expect from ELG conformity. Similarly, 1-halo conformity at high redshit was also observed in the UKIRT Infrared Deep Sky Survey \citep[UKIDSS;][]{2007Lawrence, 2015Hartley, 2016Kawinwanichakij}. 

\subsection{Origin of conformity}

A key motivation of this paper is to postulate merger-triggered starburst as a new mechanism for conformity. 
Another piece of evidence supporting this model comes in \cite{2022mHadzhiyska}, where the authors studied the clustering of mock $z = 1$ ELG and LRGs in \textsc{MilleniumTNG}. The study found strong signatures of 1-halo conformity, i.e., that ELG satellites preferentially occupy halos with ELG centrals. The study also found evidence for anisotropic distribution of ELG satellites within halos, specifically that ELG satellties preferentially exist in close proximity with each other. The authors argued that a ``cooperative star formation'' scenario can naturally explain the conformity signal seen in \textsc{MilleniumTNG}, which is consistent with our picture of merger-triggered starburst. 

Prior to this study, conformity is widely believed to originate from spatially correlated halo accretion. \cite{2016bHearin} found clear spatial correlation in halo accretion in simulations and postulated that the conformity up to $4h^{-1}$Mpc seen in SDSS is a result of SFR being correlated with halo accretion. The same work proposed that 1-halo conformity is simply a result of 2-halo conformity at higher redshift. However, this hypothesis also showed that halo accretion correlation also decreases precipitously at higher redshift, thus it would struggle to explain the strong 1-halo conformity we see at $z \sim 1$. This echoes our finding in section~\ref{sec:um} that tight correlation of SFR to halo accretion rate does not generate sufficient conformity to explain data. 

Several other studies attempted to explain 1-halo conformity via direct interactions between central and satellite galaxies. \cite{2015Hartley} studied the conformity signal between quiescent galaxies and argued that a star formation (or active galactic nucleus) related outburst event from the central galaxy could establish a hot halo environment which is then capable of quenching both central and satellite galaxies. \cite{2012Wang} similarly argued that massive red centrals preferentially live in halos with excess hot gas and massive central black holes, making them more efficient at quenching satellite galaxies. There have also been a series of papers studying mechanisms for the more contested 2-halo conformity effect, including pre-heating by early generations of black holes \citep{2015Kauffmann}, a combination of systematic biases \citep{2017Sin}, and modified halo mass quenching model \citep{2018Zu}. These mechanisms might prove relevant for 1-halo conformity as well, but we leave that to a future study. 

\subsection{Implications on merger--SFR connection}
Galaxy mergers and tidal interactions form a key part of our understanding of how galaxies form and grow. Yet there exists an active debate as to whether or how mergers trigger star formation. While studies such as \cite{2013Ellison, 2015Knapen} found mergers to suppress star formation, others found mergers to trigger periods of rapid star formation and even starbursts \citep[e.g.][]{2009Saitoh, 2021Horstman}. More recently, \cite{2022Renaud} suggested that the merger--star formation connection is dependent on the stage of evolution and can only occur after the formation of the galactic disk. Our analysis uniquely contributes to this discussion by combining information from spectra, imaging, and clustering. Our findings with DESI ELGs at $z \sim 1$ are consistent with mergers triggering star formation as ELGs exhibit rapid star formation and also tend to live in close pairs and show disturbed morphology. Our results are also qualitatively consistent with \cite{2022Renaud}'s proposal as galactic discs should have formed at redshift $z > 2$ \citep{2022Segovia}.

Ultimately, mergers and star formation are both complex processes notoriously hard to simulate. Any triggered star formation most likely depend on a plethora of properties including those of the central black holes, the molecular gas, and dynamics. DESI provides a unique window in exploring some of these relationships through its large sample size, high fidelity spectra, and broad redshift coverage. Further analyses combining DESI data and external observations in optical and radio are needed to substantiate what our findings in this paper.




\section{Conclusions}
\label{sec:conclude}

ELGs are an essential cosmological tracer as we push to higher redshifts. It is imperative to understand these galaxies so that we can ensure our cosmology pipelines are robust to potential systematics in modeling ELG bias. In this paper, we improve our understanding for the ELGs and argue for the inclusion of conformity by leveraging a variety of observations and galaxy models. 

First, we examine the inferred properties and the observed morphology of ELGs in COSMOS field, and we propose a physical picture where ELGs are starburst galaxies and tend to exhibit perturbed morphology. We postulate that at least a significant fraction of ELGs are undergoing mergers, which trigger star formation through gas inflow and tidal compression. This picture of merger-induced star formation also naturally gives rise to 1-halo conformity. 

Next we find clear evidence for 1-halo conformity in the clustering of mock ELGs and LRGs in \tng. We also summarize existing evidence for galaxy assembly bias for ELG samples. We propose an extended ELG HOD that adopts the existing skew Gaussian+power law form, but also includes galaxy assembly bias and a flexible model for 1-halo conformity. 

Then, we analyze the DESI auto and cross-correlations of the LRG and ELG samples in search of evidence of conformity in the data. We find that the DESI auto and cross-correlations favor both the inclusion of 1-halo conformity and shear-based assembly bias for the ELG HOD. However, the model preference is not conclusive based on the clustering measurements alone and we continue to find a degeneracy between small $\alpha$ and conformity.

In order to break this degeneracy, we directly constrain ELG satellite power index $\alpha$ by counting ELG satellites around redMaPPer clusters where we have estimates of halo mass from previous clustering and lensing analyses. We find that a rough $\alpha$ range of $0.8 < \alpha < 1.2$, which disfavors the vanilla HOD model, where $\alpha = 0.3$. We have taken a series of steps to reduce our exposure to systematics and conducted self-consistency tests to demonstrate a certain level of robustness, but a more thorough analysis is needed to quantitatively constrain $\alpha$ from cluster--ELG cross-correlations. Nevertheless, we find a clear positive correlation between ELG satellite number and cluster richness, rejecting models of negative $\alpha$.

Furthermore, we probe the origin of ELG conformity by stress testing UniverseMachine, a flexible model that relies on the assumption that star formation only depends on individual mass assembly of the halos. We show the existing UniverseMachine model may not be flexible enough to consistently predict the stellar mass function, the SFR distribution, and the clustering of DESI ELGs. These tests also elucidate the distinction between 1-halo conformity due to triggered starburst and conformity as an emergent phenomenon of galaxy assembly bias. We argue that an extended model for star formation that introduces a secondary dependence on the halo environment is potentially needed to consistently explain all the ELG observables. Finally, we compare our studies to existing discussions of conformity and weigh in on the debate on merger--SFR connection. 

While we have offered several pieces of evidence for ELG conformity and taken the first steps in understanding the nature of ELG, higher resolution spectra and detailed SED modeling are needed to better constrain the relevant galaxy properties. Additional observations that reveal detailed morphology and galaxy kinematics would also greatly help us confirm and dissect merger processes. In terms of HOD modeling, while our extended model can produce reasonably good fits, further refinements are likely needed to fully describe the observed clustering. These works are necessary not just in advancing our understanding of galaxy evolution beyond $z\sim 1$, but also in enabling robust cosmology with high-redshift large-scale structure analyses with the next generation of cosmological surveys. 

\section*{Acknowledgements}

We thank Wren Suess and Allison Strom for useful conversations. We thank Andrew Hearin and Yipeng Jing for constructive feedback. This work was supported by grant DE-SC0013718 and under DE-AC02-76SF00515 to SLAC National Accelerator Laboratory, and by the Kavli Institute for Particle Astrophysics and Cosmology. MdlR was supported by a Stanford Science Fellowship at Stanford University. This work was performed in part at the Aspen Center for Physics, which is supported by National Science Foundation grant PHY-2210452.

This material is based upon work supported by the U.S. Department of Energy (DOE), Office of Science, Office of High-Energy Physics, under Contract No. DE–AC02–05CH11231, and by the National Energy Research Scientific Computing Center, a DOE Office of Science User Facility under the same contract. Additional support for DESI was provided by the U.S. National Science Foundation (NSF), Division of Astronomical Sciences under Contract No. AST-0950945 to the NSF’s National Optical-Infrared Astronomy Research Laboratory; the Science and Technology Facilities Council of the United Kingdom; the Gordon and Betty Moore Foundation; the Heising-Simons Foundation; the French Alternative Energies and Atomic Energy Commission (CEA); the National Council of Science and Technology of Mexico (CONACYT); the Ministry of Science and Innovation of Spain (MICINN), and by the DESI Member Institutions: \url{https://www.desi.lbl.gov/collaborating-institutions}. Any opinions, findings, and conclusions or recommendations expressed in this material are those of the author(s) and do not necessarily reflect the views of the U. S. National Science Foundation, the U. S. Department of Energy, or any of the listed funding agencies.

The authors are honored to be permitted to conduct scientific research on Iolkam Du’ag (Kitt Peak), a mountain with particular significance to the Tohono O’odham Nation.

This work made use of Astropy:\footnote{http://www.astropy.org} a community-developed core Python package and an ecosystem of tools and resources for astronomy \citep{astropy:2013, astropy:2018, astropy:2022}.

\section*{Data Availability}

 The simulation data are available at \url{https://abacussummit.readthedocs.io/en/latest/}. The \ahod\ code package is publicly available as a part of the \textsc{abacusutils} package at \url{https://github.com/abacusorg/abacusutils}. Example usage can be found at \url{https://abacusutils.readthedocs.io/en/latest/hod.html}.
All mock products will be made available at \url{https://data.desi.lbl.gov}.



\bibliographystyle{mnras}
\bibliography{biblio} 




\appendix



\section{Author Affiliations}
\label{sec:affiliations}
$^{1}$Kavli Institute for Particle Astrophysics and Cosmology, Stanford University, 452 Lomita Mall, Stanford, CA 94305, USA\\
$^{2}$SLAC National Accelerator Laboratory, 2575 Sand Hill Road, Menlo Park, CA  94025, USA\\
$^{3}$Department of Physics,  Stanford University, 382 Via Pueblo Rd, Stanford, CA 94305, USA\\
$^{4}$Department of Physics \& Astronomy, Amherst College, 25 East Drive, Amherst, MA 01002, USA\\
$^{5}$Department of Physics, University of California, Berkeley, CA 94720, USA \\
$^{6}$Universit\'e Paris-Saclay, CEA, Institut de recherche sur les lois Fondamentales de l'Univers, 91191, Gif-sur-Yvette, France\\
$^{7}$Miller Institute for Basic Research in Science, University of California, Berkeley, CA 94720, USA\\
$^{8}$Lawrence Berkeley National Laboratory, 1 Cyclotron Road, Berkeley, CA 94720, USA\\
$^{9}$Physics Dept., Boston University, 590 Commonwealth Avenue, Boston, MA 02215, USA\\
$^{11}$Department of Physics \& Astronomy, University College London, Gower Street, London, WC1E 6BT, UK\\
$^{12}$Institute for Computational Cosmology, Department of Physics, Durham University, South Road, Durham DH1 3LE, UK\\
$^{13}$Instituto de F\'{i}sica, Universidad Nacional Aut\'{o}noma de M\'{e}xico, Cd. de M\'{e}xico C.P. 04510, M\'{e}xico\\
$^{14}$Departamento de F\'isica, Universidad de los Andes, Cra. 1 No. 18A-10, Edificio Ip, CP 111711, Bogot\'a, Colombia\\
$^{15}$Department of Physics, The Ohio State University, 191 West Woodruff Avenue, Columbus, OH 43210, USA\\
$^{16}$Center for Cosmology and AstroParticle Physics, The Ohio State University, 191 West Woodruff Avenue, Columbus, OH 43210, USA\\
$^{17}$Institut de F\'{i}sica d’Altes Energies (IFAE), The Barcelona Institute of Science and Technology, Campus UAB, 08193 Bellaterra Barcelona, Spain\\
$^{18}$NSF's NOIRLab, 950 N. Cherry Ave., Tucson, AZ 85719, USA\\
$^{19}$Instituci\'{o} Catalana de Recerca i Estudis Avan\c{c}ats, Passeig de Llu\'{\i}s Companys, 23, 08010 Barcelona, Spain\\
$^{20}$Department of Physics and Astronomy, Siena College, 515 Loudon Road, Loudonville, NY 12211, USA\\
$^{21}$National Astronomical Observatories, Chinese Academy of Sciences, A20 Datun Rd., Chaoyang District, Beijing, 100012, P.R. China\\
$^{22}$Space Sciences Laboratory, University of California, Berkeley, 7 Gauss Way, Berkeley, CA 94720, USA\\
$^{23}$Department of Physics, Kansas State University, 116 Cardwell Hall, Manhattan, KS 66506, USA\\
$^{24}$Department of Physics and Astronomy, Sejong University, Seoul, 143-747, Korea\\
$^{25}$Centro de Investigaciones Energ\'e ticas, Medioambientales y Tecnol\'o gicas (CIEMAT), Madrid, Spain
$^{26}$Department of Physics, University of Michigan, Ann Arbor, MI 48109, USA\\
$^{27}$Department of Physics and Astronomy, Ohio University, Athens, OH 45701, USA\\
$^{28}$IRFU, CEA, Universit\'{e} Paris-Saclay, F-91191 Gif-sur-Yvette, France\\

\bsp	
\label{lastpage}
\end{document}